\documentclass[aps,prl,twocolumn,groupedaddress,floatfix,showpacs]{revtex4}
\usepackage{graphicx}
\usepackage{amsmath}
\usepackage{color}

\begin{document}
\title{Atom Pairing in Optical Superlattices}

\author{J. Kangara$^{1\dagger}$, Chingyun Cheng$^{1,2\dagger}$, S. Pegahan$^1$, I. Arakelyan$^1$ and J. E. Thomas$^1$*}

\affiliation{$^{1}$Department of  Physics, North Carolina State University, Raleigh, NC 27695, USA}
\affiliation{$^{2}$Department of Physics, Duke University, Durham, NC 27708, USA}


\date{\today}

\begin{abstract}
We study the pairing of fermions in a one-dimensional lattice of tunable double-well potentials using radio-frequency spectroscopy. The spectra reveal the coexistence of two types of atom pairs with different symmetries. Our measurements  are in excellent quantitative agreement with a theoretical model, obtained by extending the Green's function method of Orso et al., [Phys. Rev. Lett. 95,  060402 (2005)], to a bichromatic 1D lattice with non-zero harmonic radial confinement. The predicted spectra comprise hundreds of discrete transitions, with symmetry-dependent initial state populations and transition strengths. Our work provides an understanding of the elementary pairing states in a superlattice, paving the way for new studies of strongly interacting many-body systems.
\end{abstract}

\maketitle

Optical superlattices, comprising two optical standing waves with a tunable relative phase,  enable wide control of the band structure of ultracold atomic gases. Ground breaking experiments with bosonic atoms in superlattices have simulated Dirac dynamics, such as Klein tunneling~\cite{WeitzKleinTunneling,WitthautDiracDynamics}, by producing linear dispersion.  A relative phase near zero creates periodic, double-well potentials with controllable asymmetry. Single atoms in the right or left states of tilted double-well potentials have been employed to study non-equilibrium dynamics~\cite{KohlNonEq} and to provide an effective spin-orbit interaction with negligible optical scattering~\cite{LiFerroSpinTexture}. This has enabled the observation of antiferromagnetic spin textures~\cite{LiFerroSpinTexture}.
Cyclic variation of the phase and corresponding double-well symmetry has been used to observe topological (Thouless) pumping for weakly interacting bosons~\cite{LohseThoulessBose} and fermions~\cite{NakajimaThoulessFermi}. Harmonic confinement, with an applied spin-dependent force produces a bilayer system, with geometric control of pairing interactions between species in separated layers~\cite{Kanasz-NagyBilayer}.  Anharmonicity in optical lattice potentials generally entangles the center of mass and relative coordinates of confinement-induced atom pairs, modifying the pair binding energy as predicted theoretically~\cite{OrsoDimerLattice} and observed in experiments~\cite{Sommer3D2D}. Anharmonic coupling also causes confinement-induced loss resonances, which  have been observed~\cite{PhysRevLett.104.153203,SalaJochimCIRexpt} and studied theoretically (see ~\cite{SalaConfinementInelastRes} and references therein), and is predicted to modify confinement induced states in a deep double-well potential~\cite{KestnerDuanDoubleWell}. With magnetically tunable two-body interactions and wide control of the dispersion relation, ultracold atomic gases in superlattices provide a broad platform for studies of many-body physics, including  entanglement, nonequilibrium dynamics, and exotic new states of matter. However, there has been no quantitative study of the elementary atom pairing states in a superlattice.

\begin{figure}[htb]
\begin{center}\
\includegraphics[width=3.25in]{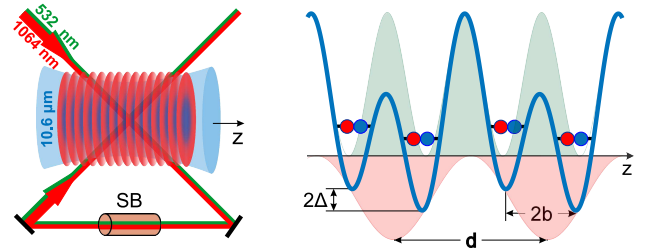}
\end{center}
\caption{A 1D optical superlattice, formed by crossed 1064/532 nm laser beams, traps atoms along z-axis, while a 10.6 $\mu$m CO$_2$ laser provides radial confinement. The potential energy for the lattice of double-wells is given by Eq.~\ref{eq:potential}  with $d$ the period and $\phi$ the relative phase set by a Soleil-Babinet compensator (SB), which determines the separation $2\,b$ and a tilt $2\Delta$ between the double-well minima. In the double-well potentials, atoms form two types of pairs with similar total energies, which can be thermally populated and probed by radio-frequency spectroscopy.
\label{fig:DimerLattice}}
\end{figure}

In this Letter, we report precision measurements of radio frequency spectra for  a 50-50 mixture of the two lowest hyperfine states (denoted $|1\rangle$, $|2\rangle$) of fermionic $^6$Li atoms in an optical superlattice, comprising attractive(red) and repulsive(green) standing waves with an adjustable relative phase. The trapped cloud is magnetically tuned near the broad collisional (Feshbach) resonance  at 832.2 G~\cite{BartensteinFeshbach,JochimPreciseFeshbach} to control the s-wave scattering length $a_{12}$. The observed spectra exhibit a rich, relative-phase dependent structure, which we explain quantitatively using a beyond Hubbard model treatment, implemented by extending the rigorous Green's function method of Orso et al.,~\cite{OrsoDimerLattice} to a 1D superlattice with non-zero harmonic radial confinement.

The bichromatic superlattice potential, Fig.~\ref{fig:DimerLattice}, is created by combining on a beam splitter two optical fields of wavelengths $\lambda_1=1064$ nm and $\lambda_2=532$ nm, with the second field obtained by frequency doubling of the first. The intensities of the two beams are controlled by acousto-optic modulators, with the green modulator operating at precisely twice the frequency of the red. The combined beams are split into two beam pairs, which intersect at an angle $\theta = 91.0^o$ to create a fundamental lattice, denoted ``red,"  with a period $d=\lambda_1/(2\sin(\theta/2))=0.75\,\mu$m and a secondary lattice, denoted ``green," with period $d/2$. The relative phase $\phi$ between the standing waves is manually tunable using a calibrated Soleil-Babinet compensator~\cite{submittedtoPRA} placed in the path of the second beam pair, to control the symmetry of the periodic double-well potential,
\begin{equation}
V(z_1) = -s_1E_R\cos^2(k z_1) + s_2E_R\cos^2(2k z_1+\phi/2),
\label{eq:potential}
\end{equation}
where $k=\pi/d$ and $E_R=\hbar^2k^2/(2m)=h\times 14.9$ kHz is the recoil energy.  A CO$_2$ laser trap propagating along the $z$-axis provides additional radial confinement. Then, $V(x_1,y_1)=m\omega_\perp^2(x_1^2+y_1^2)/2$, with $\omega_\perp=2\pi\times\beta E_R/h$ the net radial frequency and $\beta=0.0166$. Red and green lattice depths $s_1$ and $s_2$ are calibrated by modulation of the lattice amplitudes to induce inter-band transitions~\cite{submittedtoPRA}. For our experiments $s_1=7.0$, $s_2=16.5$.   The trapped cloud is typically $\simeq 30\,\mu$m in length, corresponding to $\simeq 40$ sites, with 250 atoms per site.

The atoms are cooled by evaporation near 832 G and loaded into the red lattice by increasing the intensity of the 1064 nm laser beam over 250 ms, at fixed CO$_2$ laser trap intensity. After raising the red lattice to the desired depth, the CO$_2$ laser trap is increased to provide additional radial confinement as the repulsive green lattice is ramped up over 250 ms. While the atoms are being loaded into the superlattice, the bias magnetic field is tuned to set the desired scattering length.  A radio frequency pulse of duration $\tau = 20$ ms is then applied, inducing a transitions from hyperfine state $|2\rangle$ to an initially unoccupied state $|3\rangle$. We measure the fraction of atoms lost from state $|2\rangle$ versus radio frequency $\nu$.

Spectra measured at 800.6 G probe all of the transitions from initially occupied $|12\rangle$ atom pair states with $d/a_{12}=+1.28$. The final states are $|13\rangle$ atom pair states, where $d/a_{13}=-3.78$.  For data taken in the nearly symmetric double-well configuration, $\phi \simeq 0$, we expect that two-atom states in the first and second bands will be close in energy and thermally occupied, as the single particle states are the nearly degenerate symmetric and antisymmetric states of a double-well potential,  $\varphi_\pm(z_1)\simeq[\varphi_0(z_1-b)\pm\varphi_0(z_1+b)]/\sqrt{2}$, where $\varphi_0(z_1)$ is a ground harmonic oscillator state and $2b\simeq 0.466\,d$ is the separation between the double-well minima. Shifting $\phi$ slightly away from zero localizes the center of mass in either the right or left well, strongly modifying the excitation spectra by breaking the symmetry and increasing the initial state energy separation.

\begin{figure*}[htb]
\begin{center}\
\includegraphics[width=6.0in]{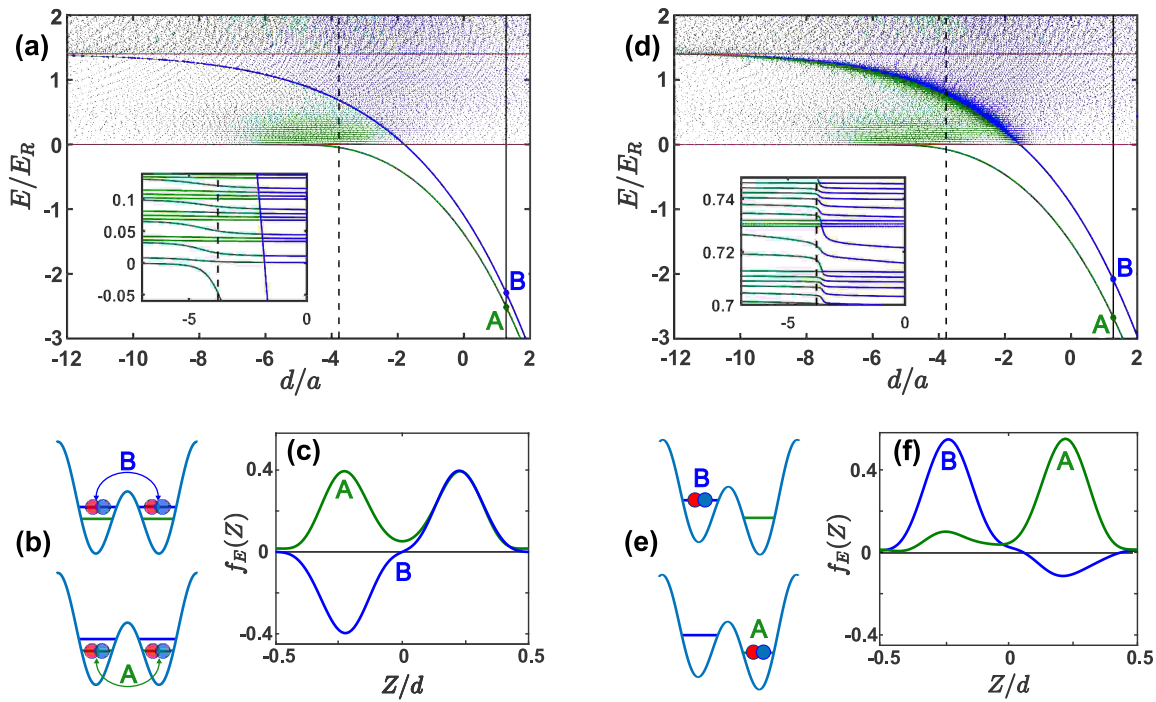}\\
\end{center}
\caption{Dimer energies $E$ for a lattice of double-well potentials versus $d/a$. For each $E$, green and blue denote the two smallest $d/a$ values. A and B show the initially populated  $|12\rangle$ dimer states with $d/a_{12}=1.28$.  Crossings with the dashed black line at $d/a_{13}=-3.78$ determine final $|13\rangle$ dimer states. The red horizontal lines denote the lowest energy for two noninteracting atoms in the first band (lower red line) and for one in each of the first two bands (upper red line). (a) Energy diagram for symmetric double wells, $\phi=0$; (b) Cartoons depict delocalized A and B dimer states. (c) Corresponding eigenstates $f_E(Z)$ (see eq.~\ref{eq:psiE}) versus  CM coordinate $Z$ are A symmetric or B antisymmetric with respect to the site center. (d) Energy diagram for tilted double wells, $\phi=\pi/35$; (e) Cartoons depict localized right (A) or left (B) dimer states. (f) Corresponding $f_E(Z)$.  Insets show typical structure for states above $E=0$.  \label{fig:Energyvsdovera}}
\end{figure*}

To understand the origin of the spectra, we begin by determining the bound eigenstates and corresponding energies for two interacting atoms in a one-dimensional bichromatic superlattice with harmonic radial confinement.  We employ a multi-band model, which is summarized briefly here and described in detail in the supplemental material~\cite{submittedtoPRA}.  Our model is based on the Green's function method of ref.~\cite{OrsoDimerLattice}, which treated the single 1D lattice case with no radial confinement. For harmonic radial confinement, the center of mass (CM) $X,Y$ motion is independent of the internal state, so we need only the energies $E$ and eigenstates for the coupled relative $\mathbf{r}\equiv(x,y,z)$ and CM $Z$ motion of the two atoms. The relevant Hamiltonian is
\begin{equation}
H(\mathbf{r},Z)=H^0(\mathbf{r},Z)+g\,\delta(\mathbf{r})\frac{\partial}{\partial r}[r...],
\label{eq:3.3}
\end{equation}
where $g=4\pi\hbar^2a_{12}/m$~\cite{effrange}  and
\begin{equation}
H^0(\mathbf{r},Z)=-\frac{\hbar^2}{2\mu}\nabla_{\mathbf{r}}^2+\frac{1}{2}\mu\omega_\perp^2r_\perp^2
-\frac{\hbar^2}{2M}\frac{\partial ^2}{\partial Z^2}+U(Z,z),
\label{eq:4.4}
\end{equation}
with $\mu=m/2$, $M=2m$ and $m$ the atom mass. Here, $r_\perp^2=x^2+y^2$ and $U(Z,z)=V(Z+z/2)+V(Z-z/2)$.

The bound state wavefunctions for an atom pair of energy $E$ and quasi-momentum $Q$ take the form
\begin{equation}
\Psi_E(\mathbf{r},Z)\propto\int dZ'G^s_E(\mathbf{r},Z;0,Z')f^{\,Q}_E(Z'),
\label{eq:psiE}
\end{equation}
where  $G^s_E$ is a Green's function, which we expand in a product basis comprising radial harmonic oscillator states and single particle Bloch states for lattice parameters $s\equiv (s_1,s_2,\phi)$. The function $f^{\,Q}_E(Z)$ is determined by solving an eigenvalue equation~\cite{submittedtoPRA}.  Using a 9-band model and 20 lattice sites, we obtain for each chosen $E$ and $Q$, 9 solutions $f^{\,Q}_E(Z)$ and corresponding $d/a$ values, arising from different combinations of CM and binding energy with the same total $E$ and $Q$.  We order the solutions by their $d/a$ values, from  most negative to most positive.

We note that $f^{\,Q}_E(Z)$ is not the CM state, as $\Psi_E(\mathbf{r},Z)$  generally does not factor, entangling the atom pair relative coordinate $\mathbf{r}$ and the CM $Z$-coordinate. However,  the  Franck-Condon factors for the transitions are proportional to the square of the overlap integrals of the $f^{\,Q}_E(Z)$ functions for the initial and final states~\cite{submittedtoPRA}, which provides substantial insight.

Figs.~\ref{fig:Energyvsdovera}(a)~and~\ref{fig:Energyvsdovera}(d) show the two lowest $d/a$ solutions for a variety of energies $E$, as green and blue dots at low resolution~\cite{submittedtoPRA}, and as continuous curves at high resolution (insets). Note that the change in color from left to right is a result of our $d/a$ labeling: For the same $E$, the smallest (left most) $d/a$ solutions are green, the next larger $d/a$ solutions are blue. For simplicity, we show predictions for $Q=0$, as the $Q$-dependence for our lattice parameters is relatively small~\cite{submittedtoPRA}. States A and B are the two bound states of lowest total energy  at $d/a_{12}=1.28$, denoted by the vertical solid black line. For symmetric double well potentials with $\phi =0$ and $Q=0$, dimer states A and B  are delocalized between the right and left wells  respectively, as depicted in Fig.~\ref{fig:Energyvsdovera}(b)  and are symmetric or antisymmetric in the CM $Z$-coordinate relative to the double-well center, as shown by the eigenstates  $f^{\,Q}_E(Z)$ of Fig.~\ref{fig:Energyvsdovera}(c). For tilted double well potentials with $\phi = \pi/35$, states A and B are localized in the right or left well, Fig.~\ref{fig:Energyvsdovera}(e),  breaking symmetry~\ref{fig:Energyvsdovera}(f) and increasing the A-B energy separation compared to Fig.~\ref{fig:Energyvsdovera}(a). The  green $E$ versus $d/a$ solid curve originating at state A asymptotes to the lowest energy of  two unbound atoms in the first band, $2\,E^1_{q_1=0}\equiv 0$, lower red horizontal line. The blue curve originating at state B asymptotes  to the lowest energy for two unbound atoms, one in each of the first and second bands, $E^2_{\pm 1}+E^1_{\mp 1}-2\,E^1_0$, upper red horizontal line.

The insets of Figs.~\ref{fig:Energyvsdovera}(a)~and~\ref{fig:Energyvsdovera}(d), for energies $E>0$, show structure similar to states studied theoretically for three dimensional harmonic confinement~\cite{Busch2HO,IdziaszekDimers}. Here, the coarse structure arises from the radial energy spacing $2\beta=0.033\,E_R$, while the finer structure arises from the lattice energy spacing, which depends on the number of sites, 20 for the model shown here~\cite{submittedtoPRA}.   For $\phi =0$ and $Q=0$, the blue curve starting at B in Fig.~\ref{fig:Energyvsdovera}(a), which arises from odd symmetry dimer states, crosses several nominally horizontal green and blue curves, which arise from even symmetry states.   In contrast, for $\phi\neq 0$, the tilted potential breaks symmetry and strongly mixes the two lowest lattice states, which have opposite symmetry. For $E>0$, this mixing changes the crossings of the blue curve in Fig.~\ref{fig:Energyvsdovera}(a) to avoided crossings in Fig.~\ref{fig:Energyvsdovera}(d), blurring the energy diagram.

\begin{figure*}[htb]
\begin{center}\
\includegraphics[width=6.0in]{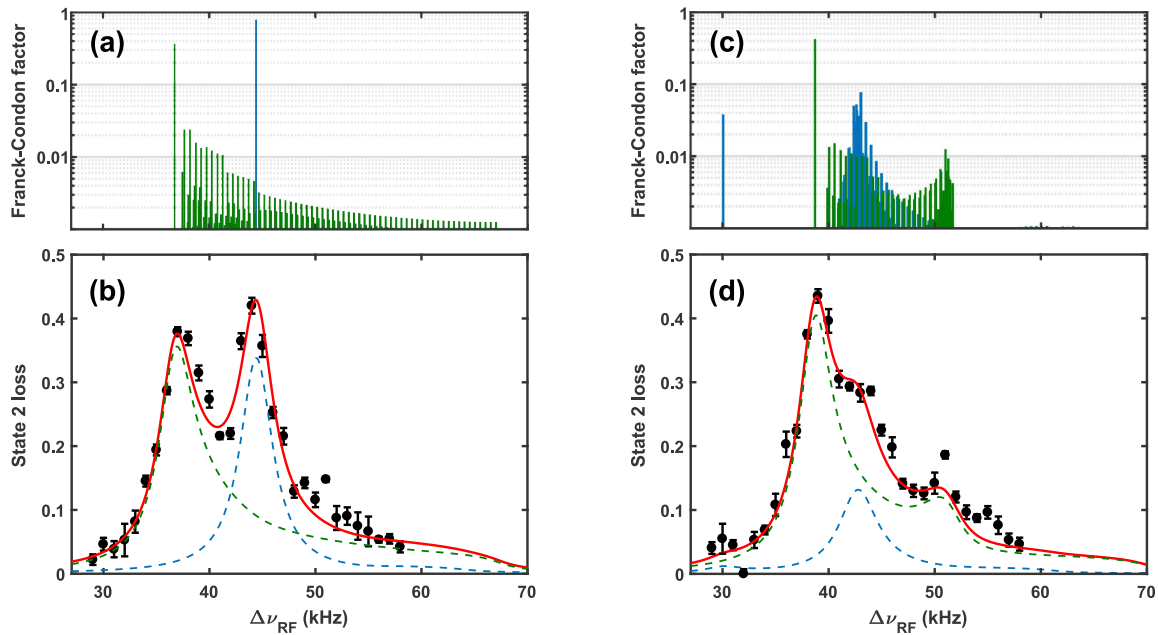}\\
\end{center}
\caption{Radio-frequency $|12\rangle\rightarrow|13\rangle$ dimer transition spectra (black dots) versus predictions (red curves). Green and blue denote contributions from states A and B of Fig.~\ref{fig:Energyvsdovera}. (a) Calculated Franck-Condon factors (log scale) for symmetric double-well potentials, $\phi =0$ as a function of transition frequency; (b) Measured spectrum showing transitions from the symmetric (A) and antisymmetric (B) states. (c) Calculated Franck-Condon factors for tilted double-well potentials, $\phi =\pi/35$; (d) Measured spectrum showing transitions from the localized right (A) and localized left (B) states of Fig.~\ref{fig:Energyvsdovera} (d). Error bars denote the standard deviation of the mean of 5 runs. \label{fig:Spectravsphi}}
\end{figure*}

To obtain the spectrum for $|12\rangle\rightarrow|13\rangle$  radio-frequency transitions, we determine the possible resonance frequencies from the energies $E_i$ of the initial pair states, where $d/a_{12}=1.28$, and the energies $E_f$ of the final pair states, where $d/a_{13}=-3.78$. The corresponding transition strengths are computed from the overlap integrals of the normalized two-atom eigenstates, $\langle f|i\rangle$. For transitions originating in dimer state $i=$ A or B, we compute the normalized spectrum,
\begin{equation}
S_i(\nu)=\frac{1}{\pi}\sum_f\frac{\gamma\,|\langle f|i\rangle|^2}{[\nu-(E_f-E_i)/h]^2+\gamma^2},
\label{eq:spectrum}
\end{equation}
where $\nu$ is the radio frequency relative to the resonance frequency of the bare atom $2\rightarrow 3$ transition. $\gamma$ denotes the spectral linewidth (HWHM) $\simeq1.8$ kHz, which is small compared to $(E_f-E_i)/h$ and comparable to that  of our previous measurements~\cite{Chingyun2DQuasi2D}.

The top panels of Figs.~\ref{fig:Spectravsphi}(a)~and~\ref{fig:Spectravsphi}(c) show the Franck-Condon factors $|\langle f|i\rangle|^2$ versus transition frequency,  for transitions from the initial bound states $i=$ A,B of Figs.~\ref{fig:Energyvsdovera}(a)~and~\ref{fig:Energyvsdovera}(d), respectively, to  final bound states $f$ with a fixed value of $d/a_{13}=-3.78$.  For $\phi=0$, transitions from the tightly bound symmetric state $A$ (green), comprise a dominant excitation to the lowest-lying, most tightly bound, symmetric state  (left peak) and to a weaker quasi-continuum of excited bound states. The latter  corresponds to a threshold spectrum for $\beta\rightarrow 0$~\cite{ZhangPolaron}. For $\phi=0$, transitions from the tightly bound antisymmetric state $B$ are dominated by a single excitation to the lowest-lying, most tightly bound, antisymmetric state (blue peak). For $\phi=\pi/35$, mixing of left- and right-well localized states increases the number of transitions from state B, blurring the spectrum near 40 kHz. Further, the lowest final state at $E<0$ acquires a non-zero overlap with the initial state B, blue peak at 30 kHz in Fig.~\ref{fig:Spectravsphi}(c). For transitions from the right-well state A, the strengths decrease quickly  above 52 kHz, as the corresponding final states become more left-well localized with increasing energy above  the fuzzy green-blue curve in Fig.~\ref{fig:Energyvsdovera}(d).

For each initial state $i=$ A or B, we find that the sum of the Franck-Condon factors, $\sum_f|\langle f|i\rangle|^2 = 0.94-0.95$, is close to unity, using only bound state solutions, eq.~\ref{eq:psiE}. This  appears to be a general property, arising from the radial confinement and periodic boundary conditions imposed on a lattice of finite length~\cite{submittedtoPRA}. Hence, we can fit the spectrum using the transition probabilities of Figs.~\ref{fig:Spectravsphi}(a)~and~\ref{fig:Spectravsphi}(c). As we expect the initial states to be thermally populated for the conditions of our experiment, we take the total spectrum to be proportional to $S(\nu)\propto\exp[-E_A/k_BT]\,S_A(\nu)+\exp[-E_B/k_BT]\,S_B(\nu)$. The red curves show the fits with $k_BT = 0.35\,E_R$ for $\phi =0$ and $0.43\,E_R$ for $\phi = \pi/35$.  An extended calculation~\cite{submittedtoPRA}, using a Boltzmann factor weighted sum over all $Q$, yields equally good fits, but with the same temperature, $k_BT=0.48\,E_R\simeq k_B\times 0.34\,\mu$K, for both $\phi=0$ and $\phi=\pi/35$.

From the very good agreement between our model and the data, we conclude that for small $\phi$, the spectra arise from two initially populated dimer states (for each $Q$), denoted $i=$A,\,B in Figs.~\ref{fig:Energyvsdovera}(a)~and~\ref{fig:Energyvsdovera}(d).
We see that the symmetry of the double-wells greatly affects both the strengths and the distribution of the transitions.

In summary, we have measured the radio-frequency spectra of  atom pair states in a 1D superlattice with radial harmonic confinement, and have developed a beyond Hubbard, multi-band  model, which explains the spectral structure. This model can be used to test the validity of analytic approximations and to  characterize the states and populations of atom pairs in general optical lattices, providing a foundation for new experiments with strongly interacting fermions.

Primary support for this research is provided by the Division of Materials Science and Engineering, the Office of Basic Energy Sciences, Office of Science, U.S. Department of Energy (DE-SC0008646). Additional support for the JETlab atom cooling group has been provided by the Physics Divisions of the Army Research Office (W911NF-14-1-0628), the National Science Foundation (PHY-1705364) and the Air Force Office of Scientific Research (FA9550-16-1-0378). \\

\noindent $^\dagger$J. K. and C. C. contributed equally to this work.

$^*$Corresponding author: jethoma7@ncsu.edu


\widetext

\appendix

\section{Supplemental Material: ``Atom Pairing in Optical Superlattices"}

In this supplemental material, we report first the methods used to calibrate the optical superlattice. Then, we describe the multi-band model employed to understand the  radio-frequency spectra for interacting atoms in a one-dimensional bichromatic lattice with nonzero radial confinement.  Resonance frequencies are calculated from the dimer binding energies. The corresponding transition strengths are determined from the overlap integrals of the two-atom eigenstates. We provide an overview of the numerical implementation of these calculations. Finally, we discuss additional spectra, which are compared to the predictions of the model including nonzero quasi-momenta.

\section{Lattice Calibration}
Calibration of the bichromatic lattice requires calibration of the relative phase and determination of the ``red" and ``green" lattice depths, which we describe below.

\subsection{Relative Phase Calibration}
\label{sec:phasecal}
The relative phase $\phi$ between the ``red" and ``green" standing waves is controlled by a micrometer on a Babinet compensator. First, we calibrate the tuning rate of the phase as a function of the micrometer reading. This is accomplished by combining the red and green beams on a beam splitter and then interfering  the beams with a small intersection angle, to create simultaneous red and green intensity standing wave patterns on a large scale. The resulting intensity profiles are imaged on a CCD array to precisely measure the change $\Delta\phi$ in the relative phase shift for a given change in the  micrometer reading.

\begin{figure*}[htb]
\begin{center}\
\includegraphics[width=6.5in]{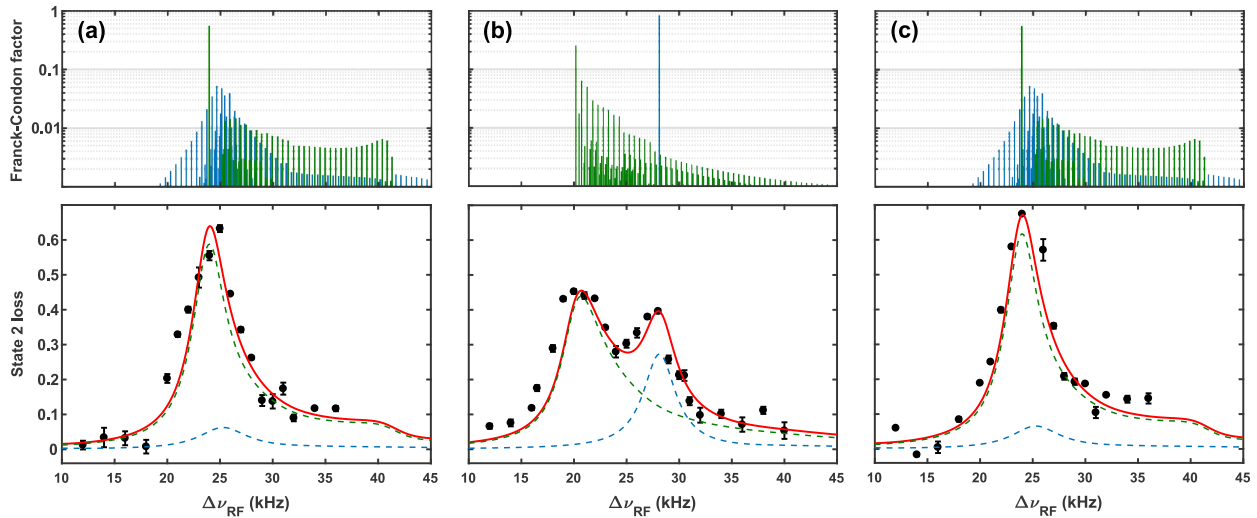}\\
\end{center}
\caption{Radio-frequency $|12\rangle\rightarrow|13\rangle$ dimer transition spectra (black dots) versus predictions (red curves) for $B=834.6$ G.  Calculated Franck-Condon factors (log scale) versus transition frequency and spectrum for (a) $\phi =-2\pi/35$; (b) $\phi =0$; (c) $\phi =+\,2\pi/35$.  Error bars denote the standard deviation of the mean of 5 runs. \label{fig:Spectra834phi}}
\end{figure*}

Next we determine the $\phi=0$ point, which is done independently of the red and green lattice depths. First, we conduct a Kapitza-Dirac scattering experiment  for various phases, using a single component gas and a pulsed the superlattice potential to imprint a spatially varying phase on the cloud. The resulting populations of negative and positive higher momentum components are unequal and interchange roles as the phase crosses either zero or $\pi$ \cite{KapizaDiracSuper2DLattice}. To distinguish the two, we measure the  radio frequency spectra of atom pairs for several phase choices: $\pi$ phase corresponds to a nominally single well potential with a higher depth and higher binding energy, than that of the 0-phase double-well. To determine the zero phase more accurately, we take spectra close to zero phase and deduce the zero point from the symmetry argument that the spectra should be identical under the change of the sign of the phase. This is illustrated in Fig.~\ref{fig:Spectra834phi} for $\phi=-2\pi/35$ and $\phi=2\pi/35$, which are symmetric about $\phi=0$. This procedure is not practical for use on a daily basis, since it takes a long time to implement.

Instead we find that the faster procedure of measuring the number of atoms loaded into the superlattice also determines the phase. For a weak CO$_2$ laser trap, the number of atoms loaded is sensitive to the radial confinement provided by the superlattice potential, since the radial confinement arising from the red and green components of the superlattice nearly cancels close to zero phase.  Shifting the phase away from zero in either direction increases loading and allows determination of the zero-phase point to better than $\pi/70$.  We verify the location of the zero phase point both before and after taking each data set. The phase $\phi$ determined by these calibration procedures is used as an input to the theoretical model described below, without further adjustment.

\subsection{Lattice Depth Measurement}
\label{sec:latticecal}
We calibrate the depth $s_1$ of the red lattice potential at 90\% of maximum power by modulation of the lattice amplitude to induce interband transitions in a single-component gas, yielding $s_1 = 15$ in recoil energy units. The results are consistent with Kapitza-Dirac scattering measurements to within 5\%, where the lattice potential is applied for a short time, imprinting a phase variation across a trapped atomic sample. Releasing atoms leads to a multi-order interference pattern, with relative contrast of the orders set by the lattice depth. We verify that the measured value of $s_1$ is consistent with the calculated depth using the measured beam powers and radii. In the spectroscopy experiments, we reduce the laser power to scale the trap depth to the chosen value of $s_1 = 7.0$.

Calibrating the green lattice using the same techniques  is more difficult, because the recoil energy is 4 times larger than that of the red.  In this case, the maximum available green lattice depth is too low for a reliable Kapitza-Dirac scattering calibration due to fast dephasing. Using lattice modulation spectroscopy at 90\% of maximum green power, we find $s_2=16$ in red recoil energy units to better than 10\% accuracy, by employing a model fit to the measured modulation spectrum. The resolution is limited by the curvature of the second and the third bands. To fit the measured radio-frequency spectra using the theoretical model described below, first we fix the red lattice depth $s_1$ and the phase $\phi$ to the calibrated values, then we adjust $s_2$. Compared to the value of $s_2=16$ measured by modulation spectroscopy for the green lattice depth, we find that $s_2 = 16.5$ gives better fits to all of the spectra, Figs.~\ref{fig:Spectra834phi}~and~\ref{fig:Spectra800phi}, which are  obtained for several different phases and $d/a$ values. Adjustment of $s_2$ by $0.5\,E_R$ produces only a small change in the peak positions. For example, near $s_2=16.5$, the lower energy peak of the spectrum in Figs.~\ref{fig:Spectra800phi}(a) varies linearly with $s_2$ with a slope of 2.3 kHz/$E_R$.

\section{Multi-Band Model of Dimer Eigenstates}
\label{sec:dimer}
To determine the eigenstates and binding energies for two atoms in a 1D bichromatic optical lattice with nonzero radial confinement, we build upon the general method of Orso, Pitaevski, Stringari, and Wooters~\cite{OrsoDimerLattice}. The required dimer wavefunctions  are the bound state solutions of the two-atom Schr\"{o}dinger equation
\begin{equation}
H\Psi(\mathbf{R},\mathbf{r})=E_{\,tot}\Psi(\mathbf{R},\mathbf{r}),
\label{eq:3.5}
\end{equation}
where $\mathbf{R}=(\mathbf{r}_1+\mathbf{r}_2)/2$ is the position of the center of mass (CM), $\mathbf{r}= \mathbf{r}_1-\mathbf{r}_2$ is the relative coordinate and $E_{\,tot}$ is the total CM and binding energy of the dimer.

The Hamiltonian is given by
\begin{equation}
H(\mathbf{R},\mathbf{r})=H_0(\mathbf{R},\mathbf{r})+g\,\delta(\mathbf{r})\frac{\partial}{\partial r}[r...],
\label{eq:3.3}
\end{equation}
where $H_0$ is the Hamiltonian for two noninteracting atoms in the optical potential and $g\equiv 4\pi\hbar^2 a/m$ determines the strength of the s-wave pseudo-potential~\cite{HuangPseudoPotential,ZwergerReview}, with $m$ the atom mass and $a$ the zero-energy scattering length. Here we have assumed that the effective range is negligible, as is the case for $^6$Li near the broad Feshbach resonances.

For a single atom, the trapping potential energy is taken to be
\begin{equation}
V(x_1,y_1,z_1)=V_\perp(x_1,y_1)+V(z_1).
\label{eq:1.4}
\end{equation}
We assume that the radial confining potential energy  is harmonic and cylindrically symmetric, $V_\perp(x_1,y_1)=\frac{1}{2}m\omega_\perp^2(x_1^2+y_1^2)$.

The axial potential energy $V(z_1)$ in the bichromatic lattice arises from two optical standing waves, a primary attractive lattice denoted ``red" and a secondary repulsive lattice, denoted ``green," as described in the main text. For the red standing wave, the periodic potential is $V_{red}(z_1)=-s_1\,E_R\,\cos^2(k z_1)$, where $E_R=\hbar^2 k^2/(2m)$ is the recoil energy, with $k=2\pi/\lambda$ the optical wavevector. Here,  $\lambda=\lambda_{red}/\sin(\theta/2)$ is the effective wavelength for two beams that intersect at an angle $\theta$. Taking the red lattice as the fundamental, $k=\pi/d$, where $d=\lambda/2$ is the lattice spacing. The  green lattice beams copropagate with the red beams and are created by frequency doubling of a portion of the red laser intensity. Hence, the effective wavelength for the green standing wave is precisely $\lambda/2$ and $V_{green}(z_1)=s_2\,E_R\,\cos^2(2k z_1+\phi/2)$. The bichromatic lattice potential for one atom is then
\begin{equation}
V(z_1)=-\frac{s_1 E_R}{2}\,\cos(G_0z_1)+\frac{s_2 E_R}{2}\,\cos(2\,G_0z_1+\phi),
\label{eq:1.1}
\end{equation}
where we have defined the fundamental reciprocal lattice vector $G_0=2k= 2\pi/d$ and eliminated the spatially constant terms.
In the experiments, the stable relative phase $\phi$ between the green and red standing wave intensities is adjusted using a Soleil-Babinet compensator for static control, calibrated as described in \S~\ref{sec:phasecal}.

For later use, we define the single particle Bloch states, which are determined from the 1D Schr\"{o}dinger equation,
\begin{equation}
\left[-\frac{\hbar^2}{2m}\frac{\partial^2}{\partial z_1^2}+V(z_1)\right]\psi^{\alpha_1}_{q_1}(z_1)=E^{\alpha_1}(q_1)\,\psi^{\alpha_1}_{q_1}(z_1),
\end{equation}
where $\alpha_1$ denotes the band and $q_1$  denotes the quasi-momentum, $-\pi/d\leq q_1\leq\pi/d$ for the first Brillouin zone. The eigenstates are given by
\begin{equation}
\psi^{\alpha_1}_{q_1}(z_1)=\sum_{G_1}C_{G_1}^{\alpha_1}(q_1)\frac{e^{i(q_1+G_1)z_1}}{\sqrt{Nd}},
\label{eq:19.2}
\end{equation}
where $G_1=(0,\pm 1,\pm 2,...)\,G_0$ is a reciprocal lattice vector and $N$ is the number of lattice sites. The states are complete on the lattice interval $0\leq z_1\leq Nd$,
\begin{equation}
\sum_{\alpha_1,q_1}\psi^{\alpha_1}_{q_1}(z_1)\psi^{\alpha_1*}_{q_1}(z_1')=\delta(z_1-z_1').
\label{eq:13.1}
\end{equation}

For two atoms, the total lattice potential is
\begin{equation}
U(z_1,z_2)=V(z_1)+V(z_2)=V(Z+z/2)+V(Z-z/2)\equiv U(Z,z).
\label{eq:2.3}
\end{equation}
As noted in ref.~\cite{OrsoDimerLattice}, we see that the CM $Z$ and relative $z$ coordinates are generally entangled by the lattice potential.

With a harmonic radial potential, for two atoms of equal mass, the CM $X,Y$ and relative $x,y$ motions  are separable, i.e., with  $M=2m$ the dimer total mass and $\mu=m/2$ the reduced mass,
\begin{equation}
U_\perp(\rho_1,\rho_2)\rightarrow U_\perp(R_\perp,r_\perp)=\frac{M\omega_\perp^2}{2}(X^2+Y^2)+\frac{\mu\omega_\perp^2}{2}(x^2+y^2),
\label{eq:2.6}
\end{equation}
Hence, we can take
\begin{equation}
\Psi(\mathbf{R},\mathbf{r})=\Phi_{CM}(X,Y)\,\Psi(\mathbf{r},Z),
\label{eq:4.1}
\end{equation}
where $\Phi_{CM}(X,Y)$ is just the harmonic oscillator state of the CM in the X-Y plane,
\begin{equation}
\left[-\frac{\hbar^2}{2M}\left(\frac{\partial^2}{\partial X^2}+\frac{\partial^2}{\partial Y^2}\right)+\frac{M\omega_\perp^2}{2}(X^2+Y^2)\right]\Phi_{CM}(X,Y)=E_{CM}^\perp\Phi_{CM}(X,Y).
\label{eq:4.2}
\end{equation}
As the orthornormal CM states $\Phi_{CM}(X,Y)$ factor out, are not coupled by the interaction, and do not change in radio frequency transitions, we will not consider them further.

The nontrivial part of the wavefunction entangles $\mathbf{r}$ and $Z$, and satisfies
\begin{equation}
\left[H^0(\mathbf{r},Z)-E\right]\Psi(\mathbf{r},Z)=-g\,\delta(\mathbf{r})\frac{\partial}{\partial r}\left[r\Psi(\mathbf{r},Z)\right],
\label{eq:5.4}
\end{equation}
where we have defined the total energy in eq.~\ref{eq:3.5} to be $E_{\,tot}=E_{CM}^\perp+E$ and
\begin{equation}
H^0(\mathbf{r},Z)=-\frac{\hbar^2}{2\mu}\nabla_{r_\perp}^2+\frac{1}{2}\mu\omega_\perp^2r_\perp^2-\frac{\hbar^2}{2\mu}\frac{\partial ^2}{\partial z^2}
-\frac{\hbar^2}{2M}\frac{\partial ^2}{\partial Z^2}+U(Z,z),
\label{eq:4.4}
\end{equation}
with $r_\perp^2=x^2+y^2$.

\subsection{Green's Function Solution}

Following ref.~\cite{OrsoDimerLattice}, we solve eq.~\ref{eq:5.4} using a Green's function method, with
\begin{equation}
\left[H^0(\mathbf{r},Z)-E\right]G_E(\mathbf{r},Z;\mathbf{r}',Z')=-\delta(\mathbf{r}-\mathbf{r}')\delta(Z-Z').
\label{eq:5.5}
\end{equation}
The formal solution to eq.~\ref{eq:5.4} for a state of energy E is then
\begin{equation}
\Psi_E(\mathbf{r},Z)=\psi_E^0(\mathbf{r},Z)+g\int dZ'\int d^3\mathbf{r}'G_E(\mathbf{r},Z;\mathbf{r}',Z')\,\delta(\mathbf{r}')\frac{\partial}{\partial r'}\left[r'\Psi_E(\mathbf{r}',Z')\right],
\label{eq:6.1}
\end{equation}
where the homogeneous solution obeys $\left[H^0(\mathbf{r},Z)-E\right]\psi_E^0(\mathbf{r},Z)=0$.

The Green's function is given in terms of a complete set of homogenous solutions satisfying $H^0(\mathbf{r},Z)\psi_\alpha(\mathbf{r},Z)=E_\alpha\,\psi_\alpha(\mathbf{r},Z)$,
\begin{equation}
\sum_\alpha\psi_\alpha(\mathbf{r},Z)\psi^*_\alpha(\mathbf{r}',Z')=\delta(\mathbf{r}-\mathbf{r}')\delta(Z-Z').
\label{eq:6.4}
\end{equation}
Then,
\begin{equation}
G_E(\mathbf{r},Z;\mathbf{r}',Z')=\sum_\alpha\frac{\psi_\alpha(\mathbf{r},Z)\psi^*_\alpha(\mathbf{r}',Z')}{E-E_\alpha+i 0^+}
\label{eq:6.5}
\end{equation}
satisfies eq.~\ref{eq:5.5}.

We are interested in the bound state solutions of eq.~\ref{eq:5.4}. In this case, the homogeneous solution in eq.~\ref{eq:6.1} is not needed and
\begin{equation}
\Psi_E(\mathbf{r},Z)= g\int dZ'\,G_E(\mathbf{r},Z;0,Z')\,\frac{\partial}{\partial r'}\left[r'\Psi_E(\mathbf{r}',Z')\right]_{r'\rightarrow 0}.
\label{eq:7.2}
\end{equation}
To solve eq.~\ref{eq:7.2}, we define
\begin{equation}
f_E(Z)=\frac{\partial}{\partial r}\left[r\Psi_E(\mathbf{r},Z)\right]_{r\rightarrow 0}.
\label{eq:7.4}
\end{equation}
Applying $g^{-1}\partial_r[r...]_{r\rightarrow 0}$ to the left hand side of eq.~\ref{eq:7.2} and using eq.~\ref{eq:7.4}, we obtain an integral eigenvalue equation as in ref.~\cite{OrsoDimerLattice},
\begin{equation}
\frac{1}{g}f_E(Z)=\int dZ' K_E(Z,Z')\,f_E(Z').
\label{eq:7.5}
\end{equation}
Here, the kernel is given by
\begin{equation}
K_E(Z,Z')=\frac{\partial}{\partial r}\left[r\,G_E(\mathbf{r},Z;0,Z')\right]_{r\rightarrow 0}.
\label{eq:7.6}
\end{equation}

As the lattice potential energy $U(Z,z)$ is periodic in $Z$, the normalized eigenstates,  eq.~\ref{eq:7.4}, can be assumed to take the Bloch form,
\begin{equation}
f^{\,Q}_E(Z)=\sum_{G'} B_{G'}^E(Q)\,\frac{e^{i(G'+Q)Z}}{\sqrt{Nd}},
\label{eq:8.1}
\end{equation}
where $G'$ is a reciprocal lattice vector and $Q$ is the total (CM) quasi-momentum, which is conserved.

Projecting eq.~\ref{eq:7.5} with $g=4\pi \hbar^2 a/m$ onto the the orthonormal basis, $e^{i(G+Q)Z}/\sqrt{Nd}$, and using  eq.~\ref{eq:8.1},  we obtain the matrix eigenvalue equation
\begin{equation}
\frac{d}{a}B_G^E(Q)=\sum_{G'} M_{GG'}(E,Q)\,B_{G'}^E(Q),
\label{eq:17.5}
\end{equation}
which is diagonal in $Q$. Here,
\begin{equation}
M_{GG'}(E,Q)=\frac{4\pi\hbar^2}{m\,N}\int dZ\int dZ'\,e^{-i(G+Q)Z+i(G'+Q)Z'}\,K_E(Z,Z').
\label{eq:18.1}
\end{equation}

To proceed further, we need to evaluate the kernel $K_E(Z,Z')$ in eq.~\ref{eq:7.6}. As pointed out in ref.~\cite{OrsoDimerLattice}, this is not trivial, since the Green's function diverges as $1/r$ at short distance, due to the contact form of the two-body interaction. Following ref.~\cite{OrsoDimerLattice}, to evaluate the kernel, we exploit the fact that the operator $\partial_r[r...]$ projects out the regular part of $G_E$ at $r\rightarrow 0$, since $\partial_r[r/r]=0$, so that the kernel is finite.

Consider first the kernel for an energy $E$ and a finite depth bichromatic lattice. We denote the lattice parameters by $s\equiv\{s_1,s_2,\phi\}$, and write
\begin{equation}
K_E^s(Z,Z')=\frac{\partial}{\partial r}\left[r\,G_E^s(\mathbf{r},Z;0,Z')\right]_{r\rightarrow 0}.
\label{eq:9.1}
\end{equation}
Subtracting the kernel for any other set of parameters $s_0$ and energy $E_0$ yields
\begin{equation}
K_E^s(Z,Z')-K_{E_0}^{s_0}(Z,Z')=\frac{\partial}{\partial r}\left\{r\left[\,G_E^s(\mathbf{r},Z;0,Z')-G_{E_0}^{s_0}(\mathbf{r},Z;0,Z')\right]\right\}_{r\rightarrow 0}.
\label{eq:9.2}
\end{equation}
As both Green's functions diverge as $1/r$ as $r\rightarrow 0$, the difference of the two Green's functions is regular as $r\rightarrow 0$. Hence, $\partial_r\{r[G_E^s-G_{E_0}^{s_0}]\}=r\partial_r(G_E^s-G_{E_0}^{s_0})+\partial_r[r]\,(G_E^s-G_{E_0}^{s_0})\rightarrow G_E^s-G_{E_0}^{s_0}$ as $r\rightarrow 0$. Then, we can write formally
\begin{equation}
K_E^s(Z,Z')= G_E^s(0,Z;0,Z')-G_{E_0}^{s_0}(0,Z;0,Z')+ K_{E_0}^{s_0}(Z,Z'),
\label{eq:10.2}
\end{equation}
where $K_{E_0}^{s_0}$ corresponds to $G_{E_0}^{s_0}$. As shown below, the evaluation is carried out so that difference of the Green's functions is manifestly finite as $r,r'\rightarrow 0$.

An important feature of eq.~\ref{eq:10.2} is that the kernel $K_E^s(Z,Z')$ is {\it independent} of the choice of the lattice parameters $s_0$ and the energy $E_0$. The evaluation is simplified by following ref.~\cite{OrsoDimerLattice}, and choosing $s_0$ to correspond to a zero depth lattice, where both the Green's function $G_{E_0}^{s_0=0}(0,Z;0,Z')$ and the kernel $K_{E_0}^{s_0=0}(Z,Z')$ are easily determined, as discussed further below.

We evaluate  eq.~\ref{eq:6.5} for $G_E$, using the complete set of separable eigenstates of the Hamiltonian of eq.~\ref{eq:4.4},
\begin{equation}
\psi_\alpha(\mathbf{r},Z)=\chi(r_\perp)\,\psi(z,Z).
\label{eq:11.2}
\end{equation}
The radial state satisfies
\begin{equation}
\left[-\frac{\hbar^2}{2\mu}\nabla_{r_\perp}^2+\frac{1}{2}\mu\omega_\perp^2r_\perp^2\right]\chi(r_\perp)=E_\perp\,\chi(r_\perp),
\label{eq:11.3}
\end{equation}
with the general orthonormal solutions
\begin{equation}
\chi_{n_\perp}^l(r_\perp,\phi)=e^{i l\phi}\sqrt{\frac{n_\perp!}{(n_\perp+|l|)!}}\,\frac{e^{-\frac{r_\perp^2}{4 l_\perp^2}}}{\sqrt{2 \pi l_\perp^2}}\left(\frac{r_\perp}{l_\perp\sqrt{2}}\right)^{|l|}
L_{n_\perp}^{|l|}\!\!\left(\frac{r_\perp^2}{2\, l_\perp^2}\right),
\end{equation}
where  $L_{n_\perp}^{|l|}(\rho)$ is an associated Laguerre polynomial, $l_\perp=\sqrt{\hbar/(m\omega_\perp)}$ is the harmonic oscillator length for one atom,
and
\begin{equation}
E_{n_\perp}^l=(2n_\perp+|l|+1)\hbar\omega_\perp.
\end{equation}
As $r_\perp,r_\perp'\rightarrow 0$ in determining the kernels, only the $l=0$ states contribute. Defining $n_\perp=m_r$ as  the radial quantum number, we take
\begin{equation}
\chi(r_\perp)\rightarrow\chi_{m_r}(r_\perp)=\frac{e^{-\frac{r_\perp^2}{4 l_\perp^2}}}{\sqrt{2 \pi l_\perp^2}}\,L^0_{m_r}\!\!\left(\frac{r_\perp^2}{2\, l_\perp^2}\right),
\label{eq:11.6}
\end{equation}
with $E_{m_r}^0=(2 m_r+1)\hbar\omega_\perp$. Here, the Laguerre polynomial is
\begin{equation}
L_{m_r}^0(\rho)=\sum_{k=0}^{m_r}\frac{(-\rho)^k\,m_r!}{(k!)^2(m_r-k)!}
\end{equation}
so that $L_{m_r}^0(0)=1$ is independent of $m_r$. These $l=0$ solutions are normalized so that
\begin{equation}
\int_0^\infty 2\pi\,r_{\!\perp} dr{\!_\perp}\,\chi^*_{m_r'}(r_\perp)\chi_{m_r}(r_\perp)=\delta_{m_r',m_r}.
\end{equation}

For the axial part of the solution $\psi(z,Z)$, we recall that
\begin{displaymath}
-\frac{\hbar^2}{2\mu}\frac{\partial ^2}{\partial z^2}
-\frac{\hbar^2}{2M}\frac{\partial ^2}{\partial Z^2}+U(Z,z)=-\frac{\hbar^2}{2m}\frac{\partial^2}{\partial z_1^2}+V(z_1)-\frac{\hbar^2}{2m}\frac{\partial^2}{\partial z_2^2}+V(z_2),
\end{displaymath}
where $z_1=Z+z/2$, $z_2=Z-z/2$. Then, with $\alpha\equiv\{\alpha_1,q_1,\alpha_2,q_2,m_r\}$, we take the required set of $l=0$ solutions to be
\begin{equation}
\psi_{\alpha}(\mathbf{r},Z)\equiv\chi_{m_r}(r_\perp)\,\psi_{q_1}^{\alpha_1}(z_1)\,\psi_{q_2}^{\alpha_2}(z_2).
\label{eq:11.7}
\end{equation}

The Green's function for $l=0$ is then given by eq.~\ref{eq:6.5} as
\begin{equation}
G_E(r_\perp,r_\perp',z_1,z_1',z_2,z_2')=\sum_{m_r,\alpha_1\!,q_1\!,\alpha_2\!,q_2}\frac{\chi_{m_r}(r_\perp)\chi^*_{m_r}(r_\perp')\psi_{q_1}^{\alpha_1}(z_1)\psi_{q_1}^{\alpha_1*}(z_1')
\psi_{q_2}^{\alpha_2}(z_2)\psi_{q_2}^{\alpha_2*}(z_2')}
{E-\hbar\omega_\perp(2m_r+1)-E_{\alpha_1}(q_1)-E_{\alpha_2}(q_2)+i0^+}
\label{eq:12.1}
\end{equation}

To obtain the kernel, eq.~\ref{eq:10.2}, we note that only the difference of two Green's functions appears, evaluated at $r_\perp,r_\perp'\rightarrow 0$. Taking the limit $z,z'\rightarrow 0$ later, we can write
\begin{eqnarray}
G_E^s(z,Z;z',Z')-G_{E_0}^{s_0}(z,Z;z',Z')&=&\nonumber\\
& &\hspace{-2in}-\frac{m}{4\pi \hbar^2}\sum_{\alpha_1\!,q_1\!,\alpha_2\!,q_2}\psi_{q_1}^{\alpha_1}(z_1)\psi_{q_1}^{\alpha_1*}(z_1')
\psi_{q_2}^{\alpha_2}(z_2)\psi_{q_2}^{\alpha_2*}(z_2')
\sum_{m_r}\frac{1}{m_r+\frac{E_{\alpha_1}(q_1)+E_{\alpha_2}(q_2)+\hbar\omega_\perp-E}{2\hbar\omega_\perp}}\nonumber\\
& &\hspace{-1.0in}-{\text same}\, (s=0,E=E_0).
\label{eq:12.2}
\end{eqnarray}
Here, the states $\psi_{q_1}^{\alpha_1}(z_1)$ and energies $E_{\alpha_1}(q_1)$ in the first term are evaluated for the nonzero lattice parameters $s$ and we have used eq.~\ref{eq:11.6} to obtain $\chi_{m_r}(0)\chi^*_{m_r}(0)/(2\hbar\omega_\perp)=m/(4\pi\hbar^2)$.

The sum over $m_r$ in eq.~\ref{eq:12.2} is convergent, since the the denominators in the $s$ and $s=0$ terms become identical in the limit $m_r\rightarrow\infty$ and the remaining sums over the band states are complete (eq.~\ref{eq:13.1}) and give $\delta(z_1-z_1')\delta(z_2-z_2')$ for any lattice depth. The sum over $m_r$ then can be evaluated using
\begin{equation}
\sum_{m_r=0}^\infty\left(\frac{1}{m_r+b}-\frac{1}{m_r+c}\right)=\psi^{(0)}(c)-\psi^{(0)}(b),
\label{eq:12.4}
\end{equation}
where  $\psi^{(n)}(x)\equiv (d/dx)^{n+1}ln[\,\Gamma(x)]$, i.e.,  polygamma$[n,x]$. The polygamma function is defined for all x, and diverges when $x$ is zero or a negative integer. Note that integral values of $x$ correspond to  energies $E$ that are resonant with a noninteracting two-atom states in eq.~\ref{eq:12.2}. For finite scattering length, bound states always correspond to non-integer $x$. We can choose the constant $c$ to be the same for both sums in eq.~\ref{eq:12.2}, as the corresponding constant $\psi^{(0)}(c)$ will cancel. Taking $b=[E_{\alpha_1}(q_1)+E_{\alpha_2}(q_2)+\hbar\omega_\perp-E]/(2\hbar\omega_\perp)$ in the first term, we can replace the sum over $m_r$ by
$-\psi^{(0)}(b)$. Taking $z = z' = 0$, we have
\begin{eqnarray}
G_E^s(0,Z;0,Z')-G_{E_0}^{s_0}(0,Z;0,Z')&=&\nonumber\\
& &\hspace{-2in}\frac{m}{4\pi \hbar^2}\sum_{\alpha_1\!,q_1\!,\alpha_2\!,q_2}\psi^{(0)}\left[\frac{\epsilon_{\alpha_1}(q_1)+\epsilon_{\alpha_2}(q_2)+\beta-\tilde{E}}{2\beta}\right]
\psi_{q_1}^{\alpha_1}(Z)\psi_{q_1}^{\alpha_1*}(Z')
\psi_{q_2}^{\alpha_2}(Z)\psi_{q_2}^{\alpha_2*}(Z')\nonumber\\
& &\hspace{-1.0 in}-{\text same}\, (s=0,E=E_0).
\label{eq:14.1}
\end{eqnarray}
Here, we have written all energies in recoil energy units, i.e., $E_{\alpha_1}(q_1)= \epsilon_{\alpha_1}(q_1)\,E_R$,  $\hbar\omega_\perp=\beta\,E_R$ and $E=\tilde{E}\,E_R$.

To solve the eigenvalue problem, eq.~\ref{eq:17.5}, according to eq.~\ref{eq:10.2}, we find the matrix elements eq.~\ref{eq:18.1} of eq.~\ref{eq:14.1},
\begin{equation}
M_{GG'}(E,E_0,Q)=\frac{4\pi\hbar^2}{m\,N}\int dZ\int dZ'\,e^{-i(G+Q)Z+i(G'+Q)Z'}\,[G_E^s(0,Z;0,Z')-G_{E_0}^{s_0}(0,Z;0,Z')].
\label{eq:19.1}
\end{equation}
Using eq.~\ref{eq:19.1}, we require
\begin{eqnarray}
I_1&=&\int_0^{Nd} dZ\,e^{-i(G+Q)Z}\psi_{q_1}^{\alpha_1}(Z)\psi_{q_2}^{\alpha_2}(Z)\nonumber\\
&=&\sum_{G_1,G_2}C_{G_1}^{\alpha_1}(q_1)C_{G_2}^{\alpha_2}(q_2)\int_0^{Nd} \frac{dZ}{Nd}e^{i(q_1+G_1+q_2+G_2-G-Q)Z},
\label{eq:19.3}
\end{eqnarray}
and similarly for the $Z'$ integral. Taking advantage of the periodicity, $\exp[i(G_1+G_2-G)nd]=1$, we let $\tilde{Z}=Z-nd$ and write
\begin{eqnarray}
\int_0^{Nd} \frac{dZ}{Nd}e^{i(q_1+G_1+q_2+G_2-G-Q)Z}&=&\sum_{n=0}^{N-1}\int_{nd}^{d+nd} \frac{dZ}{Nd}\,e^{i(q_1+G_1+q_2+G_2-G-Q)Z}\nonumber\\
&=&\frac{1}{N}\sum_{n=0}^{N-1}e^{i(q_1+q_2-Q)dn}\int_0^d\frac{d\tilde{Z}}{d}\,e^{i(q_1+G_1+q_2+G_2-G-Q)\tilde{Z}}.
\label{eq:20.1}
\end{eqnarray}
The first factor is a geometric series, which is unity for $q_1+q_2=Q+{\text integer}\times G_0$ and vanishes otherwise, since $q_1,q_2$, and $Q$ are all integer multiples of $2\pi/(Nd)$. Taking $q_1+q_2=Q$, the remaining integral is just $\delta_{G,G_1+G_2}$. The $Z$-integral is then
\begin{equation}
I_1=\sum_{G_1,G_2}C_{G_1}^{\alpha_1}(q_1)C_{G_2}^{\alpha_2}(q_2)\,\delta_{Q,q_1+q_2}\,\delta_{G,G_1+G_2}.
\label{eq:21.2}
\end{equation}
The corresponding $Z'$ integral is given by the complex conjugate of eq.~\ref{eq:21.2}, with $G,G_1,G_2\rightarrow G',G_1',G_2'$.

We define
\begin{equation}
M_{GG'}(E,Q)=M^s_{G,G'}(E,Q)-M^{(0)}_{G,G'}(E_0,Q)+ M^{0}_{G,G'}(E_0,Q),
\label{eq:21.6}
\end{equation}
where
\begin{eqnarray}
M^s_{G,G'}(E,Q)&=&\frac{1}{N}\sum_{q_1,\alpha_1,\alpha_2}\psi^{(0)}
\left[\frac{\epsilon_{\alpha_1}(q_1)+\epsilon_{\alpha_2}(Q-q_1)+\beta-\tilde{E}}{2\beta}\right]\nonumber\\
& &\sum_{G_1}C_{G_1}^{\alpha_1}(q_1)C_{G-G1}^{\alpha_2}(Q-q_1)\sum_{G_1'}C_{G_1'}^{\alpha_1*}(q_1)C_{G'-G_1'}^{\alpha_2*}(Q-q_1)
\label{eq:21.4}
\end{eqnarray}
and $M^{(0)}_{G,G'}(E_0,Q)$ is of the same form, evaluated for $s\rightarrow 0$ and $E\rightarrow E_0$. Note that the {\it difference} of the first two terms in eq.~\ref{eq:21.6} is convergent, i.e., for high band number $\alpha$, the Bloch states at finite lattice depth approach free particle states and the total energy becomes large compared to $E$ and $E_0$. From eq.~\ref{eq:10.2}, the last term, $M^0_{GG'}(E_0)$, is the matrix element of the zero lattice depth kernel $K^{s=0}_{E_0}(Z,Z')$, which we evaluate below.

We can simplify the evaluation of the $s=0$ term, $M^{(0)}_{G,G'}(E_0,Q)$, which contains free particle kinetic energies in the z-direction. Formally, for $s=0$, the coefficients $C_{G_1}^{\alpha_1}(q_1)$ for each $q_1$ are nonzero only for one value of $G_1$, i.e.,  for the first three bands,  $C^1_{G_1}(q_1)=\delta_{G_1,0}$, $C^2_{G_1}(q_1)=\delta_{G_1,-G_0}\theta[q_1]+\delta_{G_1,G_0}\theta[-q_1]$, $C^3_{G_1}(q_1)=\delta_{G_1,G_0}\theta[q_1]+\delta_{G_1,-G_0}\theta[-q_1]$. This requires $G_1'=G_1$ and $G'=G$ for the sums over reciprocal lattice vectors. Defining $G=\tilde{G}k$, $Q=\tilde{Q}k$, etc., and noting that the dimensionless kinetic energy for atom 1 is $\hbar^2(G_1+q_1)^2/(2mE_R)=(\tilde{G}_1+\tilde{q}_1)^2$, and similarly for atom 2, the sum over all bands and all $G_1$ then gives the simple result,
\begin{equation}
M^{(0)}_{G,G'}(\tilde{E}_0,\tilde{Q})=\delta_{G,G'}\,\frac{1}{N}\sum_{q_1,G_1}
\psi^{(0)}\left[\frac{(\tilde{G}_1+\tilde{q}_1)^2+(\tilde{G}+\tilde{Q}-\tilde{G}_1-\tilde{q}_1)^2+\beta-\tilde{E}_0}{2\beta}\right].
\label{eq:23.2}
\end{equation}

To complete the evaluation of eq.~\ref{eq:21.6}, we require the matrix elements $M^0_{GG'}(E_0,Q)$ of the zero lattice depth kernel $K^{s=0}_{E_0}(Z,Z')$, which are easily determined. We begin by noting that for $E=E_0$ and $s=0$, the first two terms of eq.~\ref{eq:21.6} cancel. As the momentum is conserved for zero lattice depth,  eq.~\ref{eq:17.5} is diagonal in $G$,
\begin{equation}
\frac{d}{a}B^{E_0}_G(Q)=M^0_{GG}(E_0)\,B^{E_0}_{G}(Q).
\label{eq:32.2a}
\end{equation}
For zero lattice depth, the value of $d/a$ is determined by the dimer binding energy and is independent of the CM energy. Hence, we can exploit the flexibility in the choice of $E_0$ in eq.~\ref{eq:21.6} (and eq.~\ref{eq:10.2}) to  define a {\it fixed} reference $d/a$,
\begin{equation}
M^0_{GG'}(E_0)=(d/a)_{\text ref}\,\delta_{G,G'},
\label{eq:32.2b}
\end{equation}
by choosing $E_0=E_0(G,Q)$ in the last two terms of eq.~\ref{eq:21.6} to be the total energy for a {\it fixed} binding energy $\epsilon^{\text ref}_b$ (see eq.~\ref{eq:32.3}). The value of $(d/a)_{\text ref}$  is then related to $\epsilon^{\text ref}_b$ (reference binding energy in units of $E_R$) by
\begin{equation}
\left(\frac{d}{a}\right)_{\text ref}=\pi\sqrt{\frac{\beta}{2}}\,I_{\text dimer}(\epsilon^{\text ref}_b/\beta),
\label{eq:32.8}
\end{equation}
where the scattering length and dimer binding energy are related by~\cite{ZhangPolaron,ZwergerReview},
\begin{equation}
\frac{l_\perp}{a}=I_{\text dimer}(\epsilon)\equiv\int_0^\infty\frac{dv}{\sqrt{4\pi v^3}}\left[1-\frac{2v}{1-e^{-2v}}e^{-\epsilon v}\right].
\end{equation}
Here, $\epsilon=E_b/\hbar\omega_\perp$ is the binding energy in units of $\hbar\omega_\perp=\beta\,E_R$. In eq.~\ref{eq:32.8}, we have used $d/l_\perp=\pi\sqrt{\frac{\beta}{2}}$.

For eq.~\ref{eq:32.2b} and eq.~\ref{eq:23.2} to be consistent, we use in eq.~\ref{eq:23.2} the energy,
\begin{equation}
\tilde{E}_0(G,Q,\epsilon^{\text ref}_b)=\frac{(\tilde{G}+\tilde{Q})^2}{2}+\beta -\epsilon^{\text ref}_b,
\label{eq:32.3}
\end{equation}
where the first term is the free particle CM energy of the dimer along the $z$-axis and $\beta$ is the radial ground state energy, both in units of $E_R$. From eq.~\ref{eq:32.3}, we see that the total energy argument in eq.~\ref{eq:23.2} can be written as $2\tilde{x}^2+ (\tilde{G}+\tilde{Q})^2/2+\beta-\tilde{E}_0=2\tilde{x}^2+\epsilon^{\text ref}_b$, where $\tilde{x}=\tilde{G}_1+\tilde{q}_1-(\tilde{G}+\tilde{Q})/2$. In the continuum limit, with  $\sum_{G_1,q_1}\rightarrow (N/2)\int_{-\infty}^\infty d\tilde{x}$, one can show that $M_{G,G}^{(0)}(\tilde{E}_{01},\tilde{Q})-M_{G,G}^{(0)}(\tilde{E}_{02},\tilde{Q})=d/a_1-d/a_2$, with $d/a_1$ and $d/a_2$ given by eq.~\ref{eq:32.8} and $\tilde{E}_{01}$ and $\tilde{E}_{02}$ given by eq.~\ref{eq:32.3} for binding energies $\epsilon^{\text ref}_{b1}$ and $\epsilon^{\text ref}_{b2}$ respectively. With eq.~\ref{eq:32.2b}, this result assures that the total matrix $M_{GG'}(E,Q)$ of eq.~\ref{eq:21.6} is independent of the choice of reference binding energy. For numerical evaluation with a finite number of bands, we choose $\epsilon^{\text ref}_b$ to be small compared to the maximum energy of the highest band.

Using eq.~\ref{eq:21.6} in  eq.~\ref{eq:17.5}, we find the eigenstates $f^{\,Q}_E(Z)$ and eigenvalues $d/a$  for a fixed $Q$ and selected total energy $E$.  In units of $E_R$, we take the total energy in eq.~\ref{eq:21.4} to be
\begin{equation}
\tilde{E}=2\,\epsilon_1(Q/2)+\beta-\epsilon_b.
\label{eq:13.4}
\end{equation}
Here, we follow ref.~\cite{OrsoDimerLattice} and define the binding energy $\epsilon_b$ relative to the energy of two noninteracting atoms in ground band, each with quasi-momentum $Q/2$. For the lowest band, with $\epsilon_b>0$, this procedure assures that the total bound state energy lies below the continuum. Negative values of $\epsilon_b$ then correspond to higher lying bound states.

\section{Wavefunctions and Transition Strengths}
\label{sec:transitionstrength}
In the experiments, we employ a mixture of the two lowest hyperfine states of $^6$Li, denoted $|1\rangle$, $|2\rangle$ and use a radio-frequency pulse to induce transitions from state $|2\rangle$ to an initially unpopulated state $|3\rangle$. For a given bias magnetic field, the s-wave scattering length for a $|1,2\rangle$ atom pair is generally different from that of the final $|1,3\rangle$ pair. To determine the Franck-Condon factors, we therefore need to compute the overlap integral between atom pair wavefunctions  with different energies and different $d/a$ values.

The atom pair wavefunctions for total energy $E$ are determined from eq.~\ref{eq:7.2}, using eq.~\ref{eq:7.4},
\begin{equation}
\Psi_E(\mathbf{r},Z)\propto\int dZ'\,G_E(\mathbf{r},Z;0,Z')\,f_E^{\,Q}(Z'),
\label{eq:24.1}
\end{equation}
where $f_E^{\,Q}(Z)$ is given by eq.~\ref{eq:8.1}.  $G_E(\mathbf{r},Z;0,Z')$ is given by eq.~\ref{eq:12.1}, with the relative coordinates, $r_\perp'=0$ and $z'=0$,
\begin{equation}
G_E(\mathbf{r},Z;0,Z')=\sum_{m_r,\alpha_1\!,q_1\!,\alpha_2\!,q_2}
\frac{\chi_{m_r}(r_\perp)\chi^*_{m_r}(0)\psi_{q_1}^{\alpha_1}(Z+z/2)
\psi_{q_2}^{\alpha_2}(Z-z/2)\psi_{q_1}^{\alpha_1*}(Z')\psi_{q_2}^{\alpha_2*}(Z')}
{E-\hbar\omega_\perp(2m_r+1)-E_{\alpha_1}(q_1)-E_{\alpha_2}(q_2)}.
\label{eq:24.2}
\end{equation}
The $Z'$ integral in eq.~\ref{eq:24.1} is evaluated in the same way as eq.~\ref{eq:20.1},
\begin{eqnarray}
C^{\alpha_1,\alpha_2}_{q_1,q_2}(E,Q)&=&\int_0^{Nd}dZ'\,\psi_{q_1}^{\alpha_1*}(Z')\psi_{q_2}^{\alpha_2*}(Z')\,f_E^Q(Z')\nonumber\\
&=&\delta_{Q,q_1+q_2}\sum_{G',G_1'}B_{G'}^E(Q)C_{G_1'}^{\alpha_1*}(q_1)C_{G'-G_1'}^{\alpha_2*}(q_2)\nonumber\\
&\equiv&\delta_{Q,q_1+q_2}\,\tilde{C}^{\alpha_1,\alpha_2}_q(E,Q),
\label{eq:24.5}
\end{eqnarray}
where we take  $q_1=Q+q/2$ and $q_2=Q-q/2$ to define a symmetrized coefficient,
\begin{equation}
\tilde{C}^{\alpha_1,\alpha_2}_q(E,Q)\equiv\sum_{G',G_1'}B_{G'}^E(Q)C_{G_1'}^{\alpha_1*}(Q/2+q)C_{G'-G_1'}^{\alpha_2*}(Q/2-q),
\label{eq:25.3}
\end{equation}
which is determined by the eigenstate amplitudes $B_{G'}^E(Q)$.

With these definitions, we take the normalized wavefunctions to be
\begin{eqnarray}
\Psi_E(\mathbf{r},Z)&=&\frac{A}{\sqrt{N}}\sum_q\sum_{\alpha_1,\alpha_2}\tilde{C}^{\alpha_1,\alpha_2}_q(E,Q)\,\psi_{Q/2+q}^{\alpha_1}(Z+z/2)
\,\psi_{Q/2-q}^{\alpha_2}(Z-z/2)\nonumber\\
& &\hspace{0.5in}\times\sum_{m_r}\frac{\chi_{m_r}(r_\perp)}
{m_r+\frac{\epsilon_{\alpha_1}(Q/2+q)+\epsilon_{\alpha_2}(Q/2-q)+\beta-\tilde{E}}{2\beta}},
\label{eq:28.4}
\end{eqnarray}
where all energies are in units of $E_R$ as above, and $A$ is a normalization constant. Although the wavefunction is formally divergent for $r_\perp =0$, it is normalizable, and can be used to compute the transition strengths.

We determine $A$ by requiring $\langle E|E\rangle=1=\int d^3\mathbf{r}\,dZ \,|\Psi_E(\mathbf{r},Z)|^2$. The radial integration is trivial, since the radial states are orthornormal. For the axial states, $dzdZ=dz_1dz_2$ and $\psi_{Q/2+q}^{\alpha_1}(Z+z/2)
\psi_{Q/2-q}^{\alpha_2}(Z-z/2)=\psi_{q_1}^{\alpha_1}(z_1)\psi_{q_2}^{\alpha_2}(z_2)$, which are also orthonormal. Then, for a dimer state of total energy $E_1$, we have
\begin{equation}
\langle E_1|E_1\rangle=1=|A_1|^2\frac{1}{N}\sum_q\sum_{\alpha_1,\alpha_2}|\tilde{C}^{\alpha_1,\alpha_2}_q(E_1,Q)|^2\psi^{(1)}
\left[\frac{\epsilon_{\alpha_1}(Q/2+q)+\epsilon_{\alpha_2}(Q/2-q)+\beta-\tilde{E}_1}{2\beta}\right],
\label{eq:28.1}
\end{equation}
where $\psi^{(1)}(x)=\sum_{m_r} (m_r+x)^{-2}$ is polygamma$[1,x]$.

The overlap integrals for two dimer states of total energies $E_1$ and $E_2$, $\langle E_2|E_1\rangle=\int d^3\mathbf{r}\,dZ \,\Psi^*_{E_2}(\mathbf{r},Z)\Psi_{E_1}(\mathbf{r},Z)$ are similarly determined,
\begin{eqnarray}
\langle E_2|E_1\rangle&=&A_2^*A_1\frac{1}{N}\sum_q\sum_{\alpha_1,\alpha_2}\tilde{C}^{\alpha_1,\alpha_2*}_q(E_2,Q)
\tilde{C}^{\alpha_1,\alpha_2}_q(E_1,Q)\nonumber\\
& &\frac{2\beta}{\tilde{E_2}-\tilde{E}_1}\left\{\psi^{(0)}
\left[\frac{\epsilon_{\alpha_1}(Q/2+q)+\epsilon_{\alpha_2}(Q/2-q)+\beta-\tilde{E}_1}{2\beta}\right]\right.\nonumber\\
& &\left.-\psi^{(0)}
\left[\frac{\epsilon_{\alpha_1}(Q/2+q)+\epsilon_{\alpha_2}(Q/2-q)+\beta-\tilde{E}_2}{2\beta}\right]\right\}.
\label{eq:27.6}
\end{eqnarray}
Here, we have used $\sum_{m_r}(m_r+b)^{-1}(m_r+c)^{-1}=[\psi^{(0)}(c)-\psi^{(0)}(b)]/(c-b)$. In the limit, $|\tilde{E}_2\rangle\rightarrow|\tilde{E}_1\rangle$, it is easy to show that eq.~\ref{eq:27.6} is equivalent to eq.~\ref{eq:28.1}.

Overlap integrals also can be computed from eq.~\ref{eq:17.5}, using the fact that $M_{GG'}(E,Q)$ of eq.~\ref{eq:21.6} is hermitian,
\begin{eqnarray}
\left(\frac{d}{a_1}-\frac{d}{a_2}\right)\sum_G B_G^{E_2*}(Q)B_G^{E_1}(Q)&=&\sum_{G,G'}B_G^{E_2*}(Q)[M_{GG'}(E_1,Q)-M_{GG'}(E_2,Q)]B_{G'}^{E_1}(Q)\nonumber\\
&=&\sum_{G,G'}B_G^{E_2*}(Q)[M^s_{GG'}(E_1,Q)-M^s_{GG'}(E_2,Q)]B_{G'}^{E_1}(Q),
\label{eq:29.6}
\end{eqnarray}
where the $s=0$ terms in eq.~\ref{eq:21.6} are independent of $E$ and cancel. Then, using eq.~\ref{eq:29.6}, with eqs.~\ref{eq:21.4},~\ref{eq:25.3},~and~\ref{eq:27.6}, it is straightforward to obtain
\begin{equation}
\langle E_2|E_1\rangle=A_2^*A_1\frac{2\beta}{\tilde{E}_2-\tilde{E}_1}\left(\frac{d}{a_1}-\frac{d}{a_2}\right)\sum_G B_G^{E_2*}(Q)B_G^{E_1}(Q).
\label{eq:31.1}
\end{equation}
Normalization, eq.~\ref{eq:28.1}, determines the amplitudes $A_1$ and $A_2$. Numerical evaluation confirms that eq.~\ref{eq:31.1} and eq.~\ref{eq:27.6} yield precisely the same results as they should.

Eq.~\ref{eq:31.1} shows that  $\langle E_2|E_1\rangle=0$  for  $d/a_1=d/a_2$ and $E_2-E_1\neq 0$,  i.e., dimer eigenstates of the same Hamiltonian with different total energies are orthogonal, as they should be. More importantly, Eq.~\ref{eq:31.1} shows that $\langle E_2|E_1\rangle=0$ for orthogonal eigenvectors $B_G^{E}(Q)$ of eq.~\ref{eq:17.5},  i.e., for orthogonal eigenstates $f^{\,Q}_{E_2}(Z)$ and $f^{\,Q}_{E_1}(Z)$ of eq.~\ref{eq:8.1}, which provides substantial insight, as the functions $f^{\,Q}_{E}(Z)$ are easily plotted, as shown in the main text.

\section{Numerical Implementation}
We numerically evaluate the sums appearing in eq.~\ref{eq:21.4} and eq.~\ref{eq:23.2}, using a lattice model with $9$ or more bands and $N=20$ or more lattice sites. In this case, it is important to remember that for each $G$, the range of the sum over $G_1$ in eq.~\ref{eq:21.4} must be restricted so that $G_2=G-G_1$ does not go out of range, and similarly for the sum over $G_1'$ for each $G'$.  The sum over $G_1$ in eq.~\ref{eq:23.2} for each $G$ must be restricted in the same way as that of eq.~\ref{eq:21.4}, so that  $M^{s\rightarrow 0}_{G,G'}(E\rightarrow E_0,Q)=M_{G,G'}^{(0)}(\tilde{E}_0,\tilde{Q})$, as verified numerically. This assures convergence of the difference of the sums as the energies $\epsilon_{\alpha_1}(q_1)$ become large compared to the dimer energy scales.
For nonzero dimer quasi-momentum $Q$, it is convenient, but not necessary, to symmetrize the sums over $q_1$ in eq.~\ref{eq:21.4} and eq.~\ref{eq:23.2}, by taking $\tilde{q}_1=\tilde{Q}/2+\tilde{q}$ and performing the sum over $\tilde{q}=2m_1/N$ for $-N/2\leq m_1 \leq N/2-1$, as done in \S~\ref{sec:transitionstrength} above.

\begin{figure}
\begin{center}\
\includegraphics[width=4.0in]{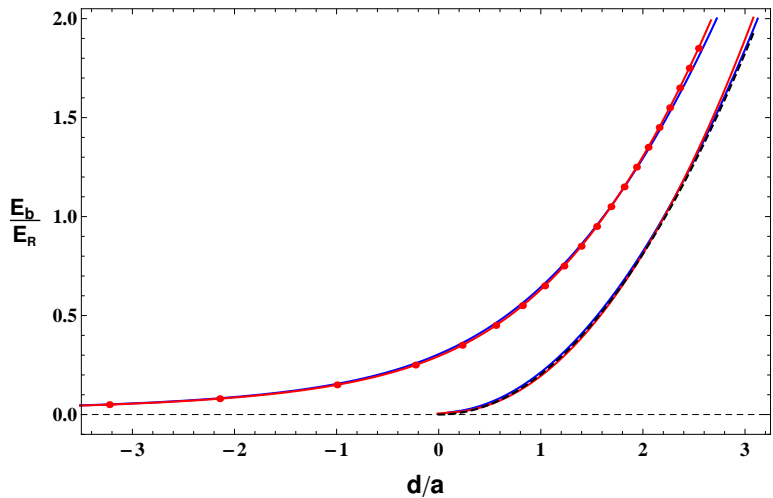}
\end{center}
\caption{Binding energy versus $d/a$ for zero lattice depth. Upper left curves for $\beta =\hbar\omega_\perp/E_R= 0.5$.  Red dots: Full diagonalization of eq.~\ref{eq:17.5} for $s=0$ (see text); Red solid curve: Analytic energy method for $s=0$ (eq.~\ref{eq:23.2}), showing exact agreement with the full diagonalization; Blue solid curve: Exact integral, eq.~\ref{eq:32.8}; Lower right curves for $\beta =\hbar\omega_\perp/E_R= 0.01$.  Red solid curve: Analytic energy method for $s=0$ (eq.~\ref{eq:23.2}); Blue solid curve: Exact integral, eq.~\ref{eq:32.8}; Black-dashed curve: Dimer binding energy in free-space, where $E_b=\hbar^2/(ma^2)$ for $a>0$, i.e., $E_b/E_R=2/\pi^2 (d/a)^2$. \label{fig:BindingZeroLattice}}
\end{figure}

To check the consistency of the numerical implementation using a fixed $(d/a)_{\text ref}$, we consider first the zero lattice depth case,  Fig.~\ref{fig:BindingZeroLattice}, for two different radial confinements, $\hbar\omega_\perp = 0.5\,E_R$, i.e., $\beta = 0.5$, and $\hbar\omega_\perp=0.01\,E_R$, which approaches the free-space limit. We initially employ a 9 band model with 20 sites and take the reference binding energy to be $\epsilon^{\text ref}_b=1.1$ in $E_R$ units, giving $(d/a)_{\text ref}=1.75$ for $\beta=0.5$ and $2.31$ for $\beta=0.01$. For $s=0$ and $\beta =0.5$, we first diagonalize eq.~\ref{eq:17.5} with $M_{G,G'}(E,Q)$ determined by eq.~\ref{eq:21.6} and $E$ by eq.~\ref{eq:13.4}.  This yields 9 different $d/a$ solutions for each input binding energy $\epsilon_b=E_b/E_R$. The lowest energy solution is displayed as the red dots on the upper left of the figure. The red solid curve shows the corresponding results with $M^s_{G,G'}(E,Q)$ replaced by a sum with the same form as eq.~\ref{eq:23.2}, and using  eq.~\ref{eq:32.3} with $\epsilon^{\text ref}_b\rightarrow\epsilon_b$. Both methods yield identical results, which are independent of $Q$, as they should be for $s=0$. The solid blue curve on the left shows the exact integral, eq.~\ref{eq:32.8}, which determines $d/a$ versus $\epsilon_b$. Shown on the lower right are the corresponding results for $\beta = 0.01$ (red solid curve) and exact integral (blue solid curve), which approach the free-space dimer binding energy (black-dashed curve), where $E_b=\hbar^2/(ma^2)$ for $a>0$, i.e., $E_b/E_R=2/\pi^2 (d/a)^2$.

For a single color lattice, and small $\beta=0.01$, we reproduce the results given for the ground band of ref.~\cite{OrsoDimerLattice}, for binding energies $\epsilon_b>0$ in $E$ of eq.~\ref{eq:13.4}. In addition, for $\epsilon_b<0$, we obtain positive energy states,  which lie above the ground state. These states are similar to those obtained for harmonic confinement in three dimensions~\cite{IdziaszekDimers}.
We also obtain additional solutions corresponding to the higher bands, which include higher lying CM states.  For a single color lattice with $s_1=2.5$ and large $\beta=10.0$, we recover the single-band Hubbard model, both numerically and analytically. In that case, for a total energy slightly below the first band two-atom  continuum,  the first solution, with the most negative $d/a$ value, corresponds to an attractive bound state. Using an energy lying above the first band two-atom  continuum (but well below the second band), the ninth solution, with the most positive $d/a$ value, corresponds to a repulsive bound state. In both cases,  the weakly bound wavefunctions are delocalized and similar in structure to those obtained by Winkler et al.,~\cite{WinklerRepulsive}.

\begin{figure}
\begin{center}\
\includegraphics[width=6.2in]{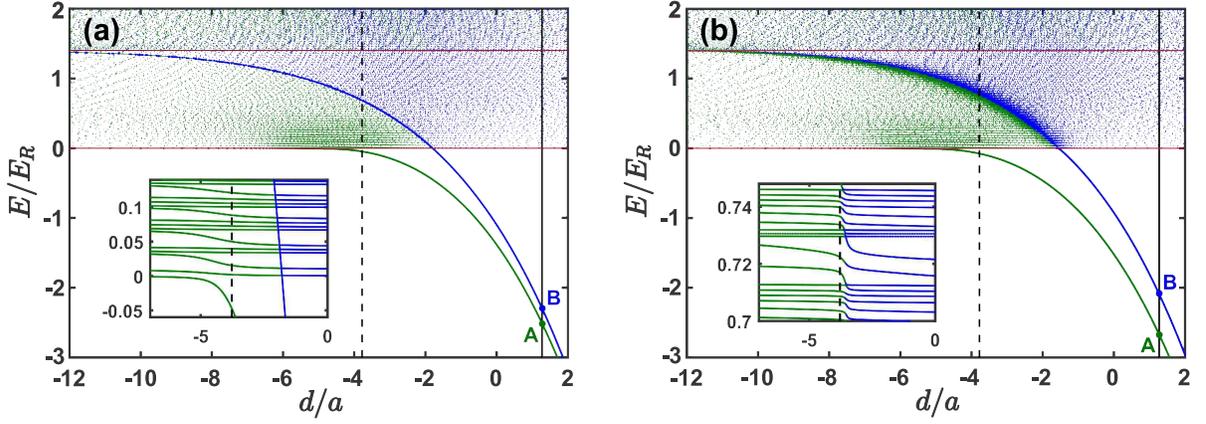}
\end{center}
\caption{Dimer energies $E$ for a lattice of double-well potentials versus $d/a$. For each $E$, green and blue denote the two smallest $d/a$ values. A and B show the initially populated  $|12\rangle$ dimer states with $d/a_{12}=1.28$.  Crossings with the dashed black line at $d/a_{13}=-3.78$ determine final $|13\rangle$ dimer states. The energy asymptotes, shown as red horizontal lines, denote the lowest energy for two noninteracting atoms in the first band (lower red line) and for one in each of the first two bands (upper red line). (a) Energy diagram for symmetric double wells, $\phi=0$;  (b) Energy diagram for tilted double wells, $\phi=\pi/35$; Insets show typical structure for states above $E=0$. \label{fig:Energyvsdovera}}
\end{figure}

\begin{figure*}[h!]
\begin{center}\
\includegraphics[width=4.7in]{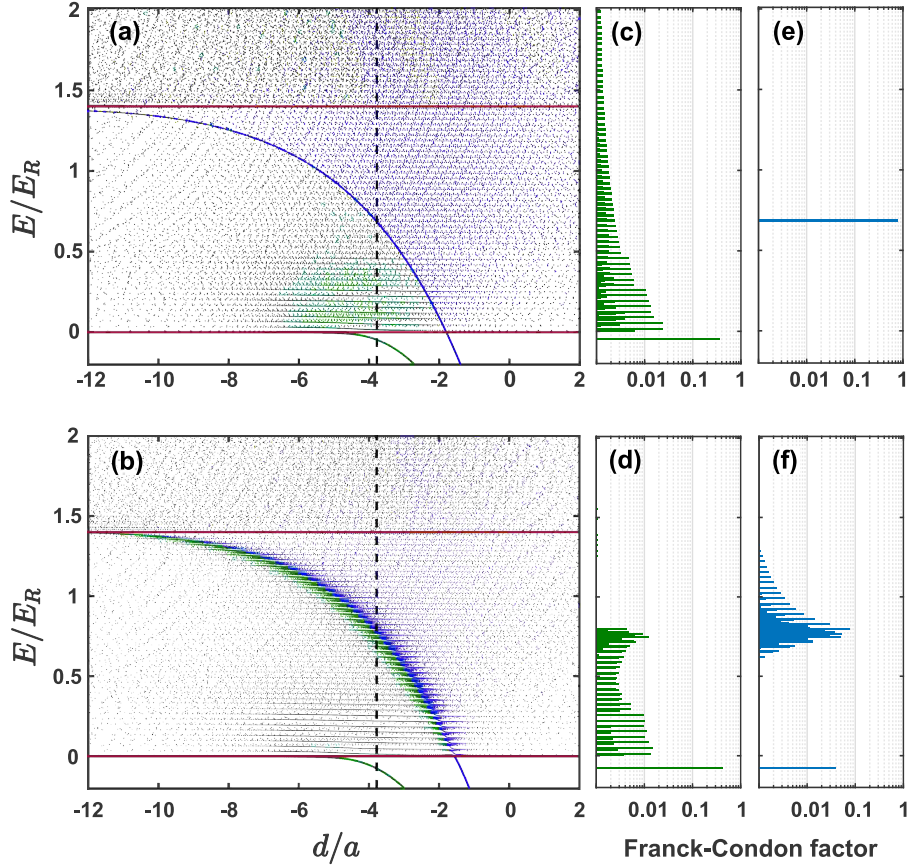}\\
\end{center}
\caption{Energies of $|13\rangle$ final states for (a) $\phi = 0$ and for (b) $\phi=\pi/35$ are determined from the crossings of the energy versus $d/a$ curves with the $d/a_{13}=-3.78$ vertical dashed black line. Panels (c-f) show the final state energies and corresponding Franck-Condon factors for transitions from the initial states A (green bars) and B (blue bars). Only the final states with
Franck-Condon factors larger than $10^{-3}$ are shown. \label{fig:FCfactors_distr}}
\end{figure*}

We find total energy $E/E_R$ versus $d/a$ curves for a variety of lattice parameters and transverse confinements. For a 9-band model, for each input energy, we obtain 9 $d/a$ solutions and order them numerically from smallest to largest, and color code, as shown in Fig.~\ref{fig:Energyvsdovera}, which is reproduced from the main paper. Only the two lowest $d/a$ solutions are plotted. The energies are input in equally spaced intervals, typically, $\Delta E=6.1\times 10^{-5}E_R$, where the interval has been decreased to the point that it contains no more than one energy value corresponding to the chosen $d/a_{13}$.  For high resolution plots,  as shown in the insets of Fig~\ref{fig:Energyvsdovera}, we employ a much smaller interval $\Delta E=6.1\times 10^{-7}E_R$, so that the $E$ versus $d/a$ curves are continuous. As noted in the main text, the coarse energy separation between the curves shown in the insets arises from the radial energy spacing, $2\beta\simeq 0.033$, for our experiment. In this case, choosing a 20 site lattice results in a lattice energy splitting smaller than the radial energy separation, producing fine structure.  Increasing the number of sites to 40 decreases this lattice energy spacing, resulting in a finer structure, and requires a smaller input energy interval to resolve the solutions. However, increasing the number of sites beyond 20 makes a negligible change in the predicted spectra. A typical energy diagram, as shown in Fig.~\ref{fig:Energyvsdovera}, can be calculated in less than 30 minutes on a personal computer with a 4-core processor.

\subsection{Evaluation of the Spectra}

Using Fig.~\ref{fig:Energyvsdovera}, we identify the set of possible final state energies from the crossings between the energy versus $d/a$ curves
(shown in detail in the insets) and the vertical dashed line corresponding to the chosen final $d/a_{13}$ value. According to eq.~\ref{eq:31.1}, the overlap integral of the initial and final states is proportional to the overlap integral of the eigenfunctions $f^{\,Q}_E(Z)$ and the normalization constants of the initial and final states. Hence, the symmetry of the $f^{\,Q}_E(Z)$ eigenstates and the localization of the wavefunctions determine the strength of the overlap integrals and hence, which identified final states can be excited. In the following, when we use the word ``states,"  we refer to the eigenstates $f^{\,Q}_E(Z)$. For $\phi = 0$,  the initial states A and B and the final states are symmetric or antisymmetric in the CM $Z$ coordinate. In this case, a transition from the antisymmetric state B  to the lowest final state at $E < 0$ is not allowed, as the two states
have opposite symmetry in $Z$. When $\phi=\pi/35$, the initial and final states can be represented as superpositions of localized right- or left-well states, and this transition is allowed.  Using eq.~\ref{eq:31.1}, we compute the squared magnitude of the overlap integrals (Franck-Condon
factors) for transitions originating from an initial $|12\rangle$ state with a given $d/a_{12}$ value to all final $|13\rangle$ states with a
fixed $d/a_{13}$ value. We find that Franck-Condon factors decrease with increasing final state energy and that the sum
over final states for each initial state converges to a value near unity.

Fig.~\ref{fig:FCfactors_distr} shows typical final state energy distributions of the Franck-Condon factors for symmetric and tilted lattices,
top and bottom rows of panels respectively. For these plots, the vertical position of each horizontal
bar corresponds to the energy of a final state. The bar lengths represent the probabilities on a log scale, where only
transitions stronger than $10^{-3}$ are shown. The green bars in panels (c) and (d) correspond to transitions from state A
of Fig. 3, while blue bars in panels (e) and (f) correspond to transitions from state B.

For $\phi = 0$, transitions from the
tightly bound lowest-lying symmetric state A, comprise a moderately strong excitation to the weakly bound, lowest-lying, symmetric final state with $E<0$ and to a
quasi-continuum of symmetric excited bound states with $E>0$ as shown in Fig.~4(c). The latter corresponds to a threshold spectrum for $\beta\rightarrow 0$~\cite{ZhangPolaron}. Transitions from the tightly bound antisymmetric state B are dominated by a strong transition to another tightly bound antisymmetric state as shown in Fig.~4(e). The binding energy and corresponding localization of the final state for B is larger than that for A, increasing the transition strength. Note that the binding energies are determined with respect to the energy asymptotes, shown as horizontal red lines in  Fig.~\ref{fig:Energyvsdovera}.  Transitions from state B to the quasi-continuum of higher lying excited bound states, above the upper energy asymptote, are weak and negligible for the measured spectrum, as the strong transition comprises most of the transition strength.

For $\phi=\pi/35$, mixing of left- and right-well localized states increases the number of possible final states for
transitions from state B, Fig.~4(f). For example, the lowest final state at $E < 0$ acquires a non-zero overlap with the
initial state B, as well as with state A, as shown in Fig.~4(f) and (d). Similarly, as seen in panel Fig.~4(f), more final
states contribute around the fuzzy border line between green and blue domains of Fig.~4(b), in contrast to the $\phi = 0$
case, where all final states except one are orthogonal to the initial state B. For transitions from the right-well state A,
Fig.~4(d), the strengths decrease quickly with increasing energy as the border line is crossed toward the blue domain,
because the final states become more left-well localized at higher energy. For transitions from the left-well state B,
Fig.~4(f), the strengths increase in the vicinity of the border line as the final states become more left-well localized
and decrease further into the blue domain due to radial delocalization of the final states.

We find that the sum of the Franck-Condon factors for transitions from a single initial bound state to all possible final bound states is always close to unity,  even for shallow lattices $s_1=2.5$ or tight radial confinement, $\beta =2.0$.  We surmise that with finite radial confinement and periodic boundary conditions for a lattice of finite length along $z$, the bound states are the only relevant solutions, i.e., formally the scattering states consist only of noninteracting states, which are orthogonal to the bound states. Similar behavior arises for simple periodic boundary conditions in a box of length $L$ in one dimension. With an interaction of the form $\alpha\,\delta (z-z_0)$ and $\alpha\neq 0$,  the formal bound state solutions obtained by the Green's function method are even in $z-z_0$ and span the space of interacting states, i.e., the solutions obtained for $\alpha=\alpha_1$ can be expanded in terms of the solutions obtained for $\alpha=\alpha_2$. In contrast, solutions which are odd in $z-z_0$ are noninteracting and irrelevant for computing Franck-Condon factors originating from an interacting state.

To predict the measured spectra, we add the contributions from all of the transitions, assuming Lorentzian lineshapes with the same width, weighted by the calculated Frank-Condon factors and centered on the  resonance frequencies corresponding to the energy differences.  For our spectral resolution, with a Lorentzian halfwidth of 1.8 kHz, we find that increasing the number of bands from 9 to 17 and the number of sites from 20 to 40 makes a negligible change in the predicted spectra.

\begin{figure*}[htb]
\begin{center}\
\includegraphics[width=6.5in]{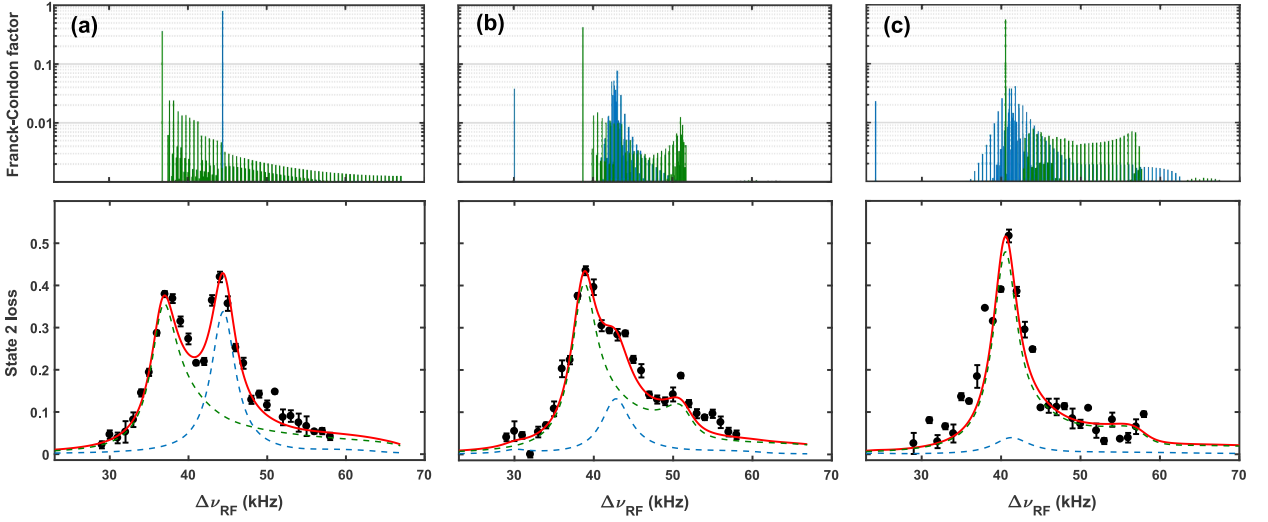}\\
\end{center}
\caption{Radio-frequency $|12\rangle\rightarrow|13\rangle$ dimer transition spectra (black dots) versus predictions (red curves) for $B=800.6$ G, $s_1=7$, $s_2=16.5$.  Calculated Franck-Condon factors (log scale) versus transition frequency and spectrum for (a) $\phi =0$; (b) $\phi =\pi/35$; (c) $\phi =2\pi/35$.  Error bars denote the standard deviation of the mean of 5 runs. \label{fig:Spectra800phi}}
\end{figure*}

Fig.~\ref{fig:Spectra800phi} compares the spectra measured at $B=800.6$ G to the model for lattice depths $s_1=7.0$ and $s_2=16.5$, determined as described in \S~\ref{sec:latticecal}. Here, we use the spectra predicted using only the $Q=0$ component, as described in the main text and further discussed in \S~\ref{sec:Q} below, where the full sum over $Q$ is determined. The red curves show the fits with $k_BT = 0.35\,E_R$ for $\phi =0$ and $0.43\,E_R$ for $\phi = \pi/35$ and $\phi=2\pi/35$. Note that the resonance frequencies are nominally twice as large as those of Fig.~\ref{fig:Spectra834phi} for $B=834.6$ G, which is fit equally well with the same parameters.

\subsection{Q-dependence of the Spectra}
\label{sec:Q}

\begin{figure*}[htb]
\begin{center}\
\includegraphics[width=6.0in]{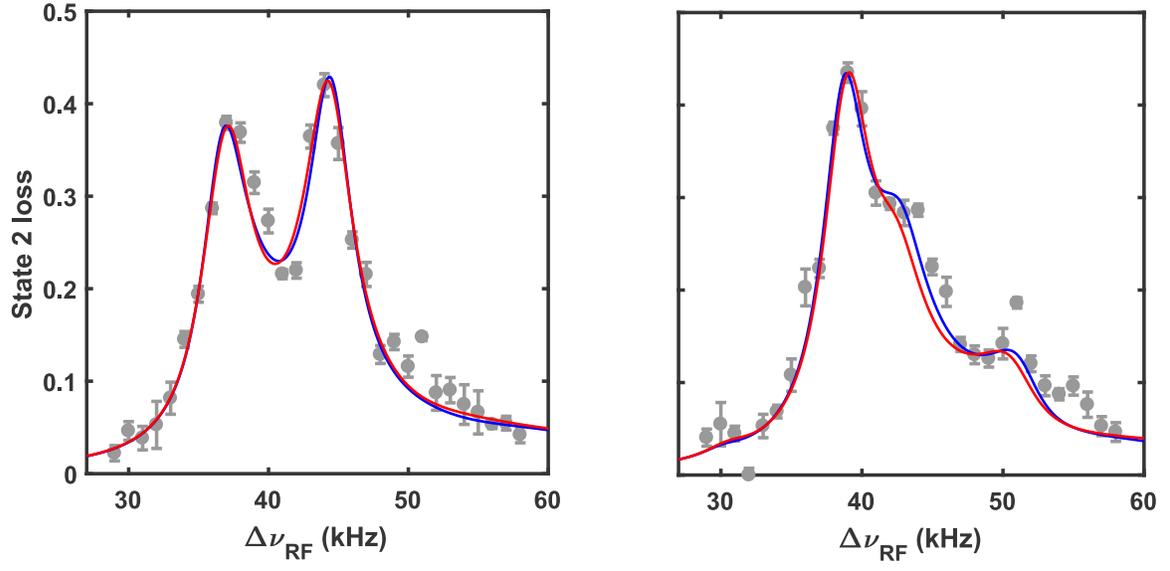}\\
\end{center}
\caption{Spectra for $B=800.6$ G calculated for $Q=0$ only (blue) versus spectra for Boltzmann-factor-weighted sum over $Q$ (red) compared to data (gray dots).  (a) $\phi =0$; (b) $\phi =\pi/35$. \label{fig:SpectraQ} }
\end{figure*}

For completeness, we consider the contribution of different Q-components to the overall spectrum. First, for each of 20 Q-values equally spaced in steps of 0.2 from -2.0 to +1.8 (one full period of the total quasi-momentum), we compute a corresponding spectrum in the same way as described above for the $Q=0$ case. Then, we weight each spectrum using a Boltzmann factor with the total energy of the corresponding Q-component given by eq.~\ref{eq:13.4} and referenced to the lowest total energy, i.e., that of two atoms in the ground band with $Q=0$, defined as $E=0$ above. Finally, we sum all of the spectral components and fit the result to the data using two parameters, the overall amplitude and a Boltzmann temperature $k_BT$. Fig.~\ref{fig:SpectraQ} compares the fits to the data for $s_1=7.0$ and $s_2=16.5$ using only the $Q=0$ component (blue) with the fit including all of the $Q$ components (red). With all of the $Q$ components included, we find that a single temperature $k_BT = 0.48E_R = k_B\times 0.34\, \mu K $ fits both the $\phi=0$ and $\phi=\pi/35$ data, in contrast to the $Q=0$ fits, where two different temperatures are required.

\begin{thebibliography}{25}
\expandafter\ifx\csname natexlab\endcsname\relax\def\natexlab#1{#1}\fi
\expandafter\ifx\csname bibnamefont\endcsname\relax
  \def\bibnamefont#1{#1}\fi
\expandafter\ifx\csname bibfnamefont\endcsname\relax
  \def\bibfnamefont#1{#1}\fi
\expandafter\ifx\csname citenamefont\endcsname\relax
  \def\citenamefont#1{#1}\fi
\expandafter\ifx\csname url\endcsname\relax
  \def\url#1{\texttt{#1}}\fi
\expandafter\ifx\csname urlprefix\endcsname\relax\def\urlprefix{URL }\fi
\providecommand{\bibinfo}[2]{#2}
\providecommand{\eprint}[2][]{\url{#2}}

\bibitem[{\citenamefont{Salger et~al.}(2011)\citenamefont{Salger, Grossert,
  Kling, and Weitz}}]{WeitzKleinTunneling}
\bibinfo{author}{\bibfnamefont{T.}~\bibnamefont{Salger}},
  \bibinfo{author}{\bibfnamefont{C.}~\bibnamefont{Grossert}},
  \bibinfo{author}{\bibfnamefont{S.}~\bibnamefont{Kling}}, \bibnamefont{and}
  \bibinfo{author}{\bibfnamefont{M.}~\bibnamefont{Weitz}},
  \bibinfo{journal}{Phys. Rev. Lett.} \textbf{\bibinfo{volume}{107}},
  \bibinfo{pages}{240401} (\bibinfo{year}{2011}).

\bibitem[{\citenamefont{Witthaut et~al.}(2011)\citenamefont{Witthaut, Salger,
  Kling, Grossert, and Weitz}}]{WitthautDiracDynamics}
\bibinfo{author}{\bibfnamefont{D.}~\bibnamefont{Witthaut}},
  \bibinfo{author}{\bibfnamefont{T.}~\bibnamefont{Salger}},
  \bibinfo{author}{\bibfnamefont{S.}~\bibnamefont{Kling}},
  \bibinfo{author}{\bibfnamefont{C.}~\bibnamefont{Grossert}}, \bibnamefont{and}
  \bibinfo{author}{\bibfnamefont{M.}~\bibnamefont{Weitz}},
  \bibinfo{journal}{Phys. Rev. A} \textbf{\bibinfo{volume}{84}},
  \bibinfo{pages}{033601} (\bibinfo{year}{2011}).

  \bibitem[{\citenamefont{Pertot et~al.}(2014)\citenamefont{Pertot, Sheikhan,
  Cocchi, Miller, Bohn, Koschorreck, K\"ohl, and Kollath}}]{KohlNonEq}
\bibinfo{author}{\bibfnamefont{D.}~\bibnamefont{Pertot}},
  \bibinfo{author}{\bibfnamefont{A.}~\bibnamefont{Sheikhan}},
  \bibinfo{author}{\bibfnamefont{E.}~\bibnamefont{Cocchi}},
  \bibinfo{author}{\bibfnamefont{L.~A.} \bibnamefont{Miller}},
  \bibinfo{author}{\bibfnamefont{J.~E.} \bibnamefont{Bohn}},
  \bibinfo{author}{\bibfnamefont{M.}~\bibnamefont{Koschorreck}},
  \bibinfo{author}{\bibfnamefont{M.}~\bibnamefont{K\"ohl}}, \bibnamefont{and}
  \bibinfo{author}{\bibfnamefont{C.}~\bibnamefont{Kollath}},
  \bibinfo{journal}{Phys. Rev. Lett.} \textbf{\bibinfo{volume}{113}},
  \bibinfo{pages}{170403} (\bibinfo{year}{2014}).

\bibitem[{\citenamefont{Li et~al.}(2016)\citenamefont{Li, Huang, Shteynas,
  Burchesky, Top, Su, Lee, Jamison, and Ketterle}}]{LiFerroSpinTexture}
\bibinfo{author}{\bibfnamefont{J.}~\bibnamefont{Li}},
  \bibinfo{author}{\bibfnamefont{W.}~\bibnamefont{Huang}},
  \bibinfo{author}{\bibfnamefont{B.}~\bibnamefont{Shteynas}},
  \bibinfo{author}{\bibfnamefont{S.}~\bibnamefont{Burchesky}},
  \bibinfo{author}{\bibfnamefont{F.~C.} \bibnamefont{Top}},
  \bibinfo{author}{\bibfnamefont{E.}~\bibnamefont{Su}},
  \bibinfo{author}{\bibfnamefont{J.}~\bibnamefont{Lee}},
  \bibinfo{author}{\bibfnamefont{A.~O.} \bibnamefont{Jamison}},
  \bibnamefont{and} \bibinfo{author}{\bibfnamefont{W.}~\bibnamefont{Ketterle}},
  \bibinfo{journal}{Phys. Rev. Lett.} \textbf{\bibinfo{volume}{117}},
  \bibinfo{pages}{185301} (\bibinfo{year}{2016}).

\bibitem[{\citenamefont{Lohse et~al.}(2016)\citenamefont{Lohse, Schweizer,
  Zilberberg, Aidelsburger, and Bloch}}]{LohseThoulessBose}
\bibinfo{author}{\bibfnamefont{M.}~\bibnamefont{Lohse}},
  \bibinfo{author}{\bibfnamefont{C.}~\bibnamefont{Schweizer}},
  \bibinfo{author}{\bibfnamefont{O.}~\bibnamefont{Zilberberg}},
  \bibinfo{author}{\bibfnamefont{M.}~\bibnamefont{Aidelsburger}},
  \bibnamefont{and} \bibinfo{author}{\bibfnamefont{I.}~\bibnamefont{Bloch}},
  \bibinfo{journal}{Nature Phys.} \textbf{\bibinfo{volume}{12}},
  \bibinfo{pages}{350} (\bibinfo{year}{2016}).

\bibitem[{\citenamefont{Nakajima et~al.}(2016)\citenamefont{Nakajima, Tomita,
  Taie, Ichinose, Ozawa, Wang, Troyer, and Takahashi}}]{NakajimaThoulessFermi}
\bibinfo{author}{\bibfnamefont{S.}~\bibnamefont{Nakajima}},
  \bibinfo{author}{\bibfnamefont{T.}~\bibnamefont{Tomita}},
  \bibinfo{author}{\bibfnamefont{S.}~\bibnamefont{Taie}},
  \bibinfo{author}{\bibfnamefont{T.}~\bibnamefont{Ichinose}},
  \bibinfo{author}{\bibfnamefont{H.}~\bibnamefont{Ozawa}},
  \bibinfo{author}{\bibfnamefont{L.}~\bibnamefont{Wang}},
  \bibinfo{author}{\bibfnamefont{M.}~\bibnamefont{Troyer}}, \bibnamefont{and}
  \bibinfo{author}{\bibfnamefont{Y.}~\bibnamefont{Takahashi}},
  \bibinfo{journal}{Nature Phys.} \textbf{\bibinfo{volume}{12}},
  \bibinfo{pages}{296} (\bibinfo{year}{2016}).

\bibitem[{\citenamefont{Kan\'{a}sz-Nagy
  et~al.}(2015)\citenamefont{Kan\'{a}sz-Nagy, Demler, and
  Zar\'{a}nd}}]{Kanasz-NagyBilayer}
\bibinfo{author}{\bibfnamefont{M.}~\bibnamefont{Kan\'{a}sz-Nagy}},
  \bibinfo{author}{\bibfnamefont{E.~A.} \bibnamefont{Demler}},
  \bibnamefont{and}
  \bibinfo{author}{\bibfnamefont{G.}~\bibnamefont{Zar\'{a}nd}},
  \bibinfo{journal}{Phys. Rev. A} \textbf{\bibinfo{volume}{91}},
  \bibinfo{pages}{032704} (\bibinfo{year}{2015}).

\bibitem[{\citenamefont{Orso et~al.}(2005)\citenamefont{Orso, Pitaevskii,
  Stringari, and Wouters}}]{OrsoDimerLattice}
\bibinfo{author}{\bibfnamefont{G.}~\bibnamefont{Orso}},
  \bibinfo{author}{\bibfnamefont{L.~P.} \bibnamefont{Pitaevskii}},
  \bibinfo{author}{\bibfnamefont{S.}~\bibnamefont{Stringari}},
  \bibnamefont{and} \bibinfo{author}{\bibfnamefont{M.}~\bibnamefont{Wouters}},
  \bibinfo{journal}{Phys. Rev. Lett.} \textbf{\bibinfo{volume}{95}},
  \bibinfo{pages}{060402} (\bibinfo{year}{2005}).

\bibitem[{\citenamefont{Sommer et~al.}(2012)\citenamefont{Sommer, Cheuk, Ku,
  Bakr, and Zwierlein}}]{Sommer3D2D}
\bibinfo{author}{\bibfnamefont{A.~T.} \bibnamefont{Sommer}},
  \bibinfo{author}{\bibfnamefont{L.~W.} \bibnamefont{Cheuk}},
  \bibinfo{author}{\bibfnamefont{M.~J.~H.} \bibnamefont{Ku}},
  \bibinfo{author}{\bibfnamefont{W.~S.} \bibnamefont{Bakr}}, \bibnamefont{and}
  \bibinfo{author}{\bibfnamefont{M.~W.} \bibnamefont{Zwierlein}},
  \bibinfo{journal}{Phys. Rev. Lett.} \textbf{\bibinfo{volume}{108}},
  \bibinfo{pages}{045302} (\bibinfo{year}{2012}).

\bibitem[{\citenamefont{Haller et~al.}(2010)\citenamefont{Haller, Mark, Hart,
  Danzl, Reichs\"ollner, Melezhik, Schmelcher, and
  N\"agerl}}]{PhysRevLett.104.153203}
\bibinfo{author}{\bibfnamefont{E.}~\bibnamefont{Haller}},
  \bibinfo{author}{\bibfnamefont{M.~J.} \bibnamefont{Mark}},
  \bibinfo{author}{\bibfnamefont{R.}~\bibnamefont{Hart}},
  \bibinfo{author}{\bibfnamefont{J.~G.} \bibnamefont{Danzl}},
  \bibinfo{author}{\bibfnamefont{L.}~\bibnamefont{Reichs\"ollner}},
  \bibinfo{author}{\bibfnamefont{V.}~\bibnamefont{Melezhik}},
  \bibinfo{author}{\bibfnamefont{P.}~\bibnamefont{Schmelcher}},
  \bibnamefont{and} \bibinfo{author}{\bibfnamefont{H.-C.}
  \bibnamefont{N\"agerl}}, \bibinfo{journal}{Phys. Rev. Lett.}
  \textbf{\bibinfo{volume}{104}}, \bibinfo{pages}{153203}
  (\bibinfo{year}{2010}).

\bibitem[{\citenamefont{Sala et~al.}(2013)\citenamefont{Sala, Z\"urn, Lompe,
  Wenz, Murmann, Serwane, Jochim, and Saenz}}]{SalaJochimCIRexpt}
\bibinfo{author}{\bibfnamefont{S.}~\bibnamefont{Sala}},
  \bibinfo{author}{\bibfnamefont{G.}~\bibnamefont{Z\"urn}},
  \bibinfo{author}{\bibfnamefont{T.}~\bibnamefont{Lompe}},
  \bibinfo{author}{\bibfnamefont{A.~N.} \bibnamefont{Wenz}},
  \bibinfo{author}{\bibfnamefont{S.}~\bibnamefont{Murmann}},
  \bibinfo{author}{\bibfnamefont{F.}~\bibnamefont{Serwane}},
  \bibinfo{author}{\bibfnamefont{S.}~\bibnamefont{Jochim}}, \bibnamefont{and}
  \bibinfo{author}{\bibfnamefont{A.}~\bibnamefont{Saenz}},
  \bibinfo{journal}{Phys. Rev. Lett.} \textbf{\bibinfo{volume}{110}},
  \bibinfo{pages}{203202} (\bibinfo{year}{2013}).

\bibitem[{\citenamefont{Sala and Saenz}(2016)}]{SalaConfinementInelastRes}
\bibinfo{author}{\bibfnamefont{S.}~\bibnamefont{Sala}} \bibnamefont{and}
  \bibinfo{author}{\bibfnamefont{A.}~\bibnamefont{Saenz}},
  \bibinfo{journal}{Phys. Rev. A} \textbf{\bibinfo{volume}{94}},
  \bibinfo{pages}{022713} (\bibinfo{year}{2016}).

\bibitem[{\citenamefont{Kester and Duan}(2012)}]{KestnerDuanDoubleWell}
\bibinfo{author}{\bibfnamefont{J.~P.} \bibnamefont{Kester}} \bibnamefont{and}
  \bibinfo{author}{\bibfnamefont{L.-M.} \bibnamefont{Duan}},
  \bibinfo{journal}{New. J. Phys.} \textbf{\bibinfo{volume}{12}},
  \bibinfo{pages}{05316} (\bibinfo{year}{2012}).

\bibitem[{\citenamefont{Bartenstein et~al.}(2005)\citenamefont{Bartenstein,
  Altmeyer, Riedl, Geursen, Jochim, Chin, Denschlag, Grimm, Simoni, Tiesinga
  et~al.}}]{BartensteinFeshbach}
\bibinfo{author}{\bibfnamefont{M.}~\bibnamefont{Bartenstein}},
  \bibinfo{author}{\bibfnamefont{A.}~\bibnamefont{Altmeyer}},
  \bibinfo{author}{\bibfnamefont{S.}~\bibnamefont{Riedl}},
  \bibinfo{author}{\bibfnamefont{R.}~\bibnamefont{Geursen}},
  \bibinfo{author}{\bibfnamefont{S.}~\bibnamefont{Jochim}},
  \bibinfo{author}{\bibfnamefont{C.}~\bibnamefont{Chin}},
  \bibinfo{author}{\bibfnamefont{J.~H.} \bibnamefont{Denschlag}},
  \bibinfo{author}{\bibfnamefont{R.}~\bibnamefont{Grimm}},
  \bibinfo{author}{\bibfnamefont{A.}~\bibnamefont{Simoni}},
  \bibinfo{author}{\bibfnamefont{E.}~\bibnamefont{Tiesinga}},
  \bibnamefont{et~al.}, \bibinfo{journal}{Phys. Rev. Lett.}
  \textbf{\bibinfo{volume}{94}}, \bibinfo{pages}{103201}
  (\bibinfo{year}{2005}).

\bibitem[{\citenamefont{Z\"urn et~al.}(2013)\citenamefont{Z\"urn, Lompe, Wenz,
  Jochim, Julienne, and Hutson}}]{JochimPreciseFeshbach}
\bibinfo{author}{\bibfnamefont{G.}~\bibnamefont{Z\"urn}},
  \bibinfo{author}{\bibfnamefont{T.}~\bibnamefont{Lompe}},
  \bibinfo{author}{\bibfnamefont{A.~N.} \bibnamefont{Wenz}},
  \bibinfo{author}{\bibfnamefont{S.}~\bibnamefont{Jochim}},
  \bibinfo{author}{\bibfnamefont{P.~S.} \bibnamefont{Julienne}},
  \bibnamefont{and} \bibinfo{author}{\bibfnamefont{J.~M.}
  \bibnamefont{Hutson}}, \bibinfo{journal}{Phys. Rev. Lett.}
  \textbf{\bibinfo{volume}{110}}, \bibinfo{pages}{135301}
  (\bibinfo{year}{2013}).

\bibitem[{sub()}]{submittedtoPRA}
\bibinfo{note}{See Supplemental Material at http://link.aps.org/supplemental/
  for details on lattice calibration, theoretical model of dimer binding in a
  superlattice, and its numerical implementation, which includes
  Refs.~\cite{KapizaDiracSuper2DLattice,HuangPseudoPotential,ZwergerReview,WinklerRepulsive}}.

\bibitem[{\citenamefont{Jo et~al.}(2012)\citenamefont{Jo, Guzman, Thomas,
  Hosur, Vishwanath, and Stamper-Kurn}}]{KapizaDiracSuper2DLattice}
\bibinfo{author}{\bibfnamefont{G.-B.} \bibnamefont{Jo}},
  \bibinfo{author}{\bibfnamefont{J.}~\bibnamefont{Guzman}},
  \bibinfo{author}{\bibfnamefont{C.~K.} \bibnamefont{Thomas}},
  \bibinfo{author}{\bibfnamefont{P.}~\bibnamefont{Hosur}},
  \bibinfo{author}{\bibfnamefont{A.}~\bibnamefont{Vishwanath}},
  \bibnamefont{and} \bibinfo{author}{\bibfnamefont{D.~M.}
  \bibnamefont{Stamper-Kurn}}, \bibinfo{journal}{Phys. Rev. Lett.}
  \textbf{\bibinfo{volume}{108}}, \bibinfo{pages}{045305}
  (\bibinfo{year}{2012}).

\bibitem[{\citenamefont{Huang}(1963)}]{HuangPseudoPotential}
\bibinfo{author}{\bibfnamefont{K.}~\bibnamefont{Huang}},
  \emph{\bibinfo{title}{Statistical Mechanics}} (\bibinfo{publisher}{Wiley},
  \bibinfo{year}{1963}), p. \bibinfo{pages}{455}.

\bibitem[{\citenamefont{Bloch et~al.}(2008)\citenamefont{Bloch, Dalibard, and
  Zwerger}}]{ZwergerReview}
\bibinfo{author}{\bibfnamefont{I.}~\bibnamefont{Bloch}},
  \bibinfo{author}{\bibfnamefont{J.}~\bibnamefont{Dalibard}}, \bibnamefont{and}
  \bibinfo{author}{\bibfnamefont{W.}~\bibnamefont{Zwerger}},
  \bibinfo{journal}{Rev. Mod. Phys.} \textbf{\bibinfo{volume}{80}},
  \bibinfo{pages}{885} (\bibinfo{year}{2008}).

\bibitem[{\citenamefont{Winkler et~al.}(2006)\citenamefont{Winkler, Thalhammer,
  Lang, Grimm, Hecker~Denschlag, Daley, Kantian, {B\"{u}chler}, and
  Zoller}}]{WinklerRepulsive}
\bibinfo{author}{\bibfnamefont{K.}~\bibnamefont{Winkler}},
  \bibinfo{author}{\bibfnamefont{G.}~\bibnamefont{Thalhammer}},
  \bibinfo{author}{\bibfnamefont{F.}~\bibnamefont{Lang}},
  \bibinfo{author}{\bibfnamefont{R.}~\bibnamefont{Grimm}},
  \bibinfo{author}{\bibfnamefont{J.}~\bibnamefont{Hecker~Denschlag}},
  \bibinfo{author}{\bibfnamefont{A.~J.} \bibnamefont{Daley}},
  \bibinfo{author}{\bibfnamefont{A.}~\bibnamefont{Kantian}},
  \bibinfo{author}{\bibfnamefont{H.~P.} \bibnamefont{{B\"{u}chler}}},
  \bibnamefont{and} \bibinfo{author}{\bibfnamefont{P.}~\bibnamefont{Zoller}},
  \bibinfo{journal}{Nature} \textbf{\bibinfo{volume}{441}},
  \bibinfo{pages}{853} (\bibinfo{year}{2006}).

\bibitem[{eff()}]{effrange}
\bibinfo{note}{We assume that the effective range is negligible, which is a
  good approximation for the broad Feshbach resonances in $^6$Li.}

\bibitem[{\citenamefont{Busch et~al.}(1998)\citenamefont{Busch, Englert,
  Rza{\.{z}}ewski, and Wilkens}}]{Busch2HO}
\bibinfo{author}{\bibfnamefont{T.}~\bibnamefont{Busch}},
  \bibinfo{author}{\bibfnamefont{B.-G.} \bibnamefont{Englert}},
  \bibinfo{author}{\bibfnamefont{K.}~\bibnamefont{Rza{\.{z}}ewski}},
  \bibnamefont{and} \bibinfo{author}{\bibfnamefont{M.}~\bibnamefont{Wilkens}},
  \bibinfo{journal}{Foundations of Physics} \textbf{\bibinfo{volume}{28}},
  \bibinfo{pages}{549} (\bibinfo{year}{1998}).

\bibitem[{\citenamefont{Idziaszek and Calarco}(2006)}]{IdziaszekDimers}
\bibinfo{author}{\bibfnamefont{Z.}~\bibnamefont{Idziaszek}} \bibnamefont{and}
  \bibinfo{author}{\bibfnamefont{T.}~\bibnamefont{Calarco}},
  \bibinfo{journal}{Phys. Rev. A} \textbf{\bibinfo{volume}{74}},
  \bibinfo{pages}{022712} (\bibinfo{year}{2006}).

\bibitem[{\citenamefont{Cheng et~al.}(2016)\citenamefont{Cheng, Kangara,
  Arakelyan, and Thomas}}]{Chingyun2DQuasi2D}
\bibinfo{author}{\bibfnamefont{C.}~\bibnamefont{Cheng}},
  \bibinfo{author}{\bibfnamefont{J.}~\bibnamefont{Kangara}},
  \bibinfo{author}{\bibfnamefont{I.}~\bibnamefont{Arakelyan}},
  \bibnamefont{and} \bibinfo{author}{\bibfnamefont{J.~E.}
  \bibnamefont{Thomas}}, \bibinfo{journal}{Phys. Rev. A}
  \textbf{\bibinfo{volume}{94}}, \bibinfo{pages}{031606}
  (\bibinfo{year}{2016}).

\bibitem[{\citenamefont{Zhang et~al.}(2012)\citenamefont{Zhang, Ong, Arakelyan,
  and Thomas}}]{ZhangPolaron}
\bibinfo{author}{\bibfnamefont{Y.}~\bibnamefont{Zhang}},
  \bibinfo{author}{\bibfnamefont{W.}~\bibnamefont{Ong}},
  \bibinfo{author}{\bibfnamefont{I.}~\bibnamefont{Arakelyan}},
  \bibnamefont{and} \bibinfo{author}{\bibfnamefont{J.~E.}
  \bibnamefont{Thomas}}, \bibinfo{journal}{Phys. Rev. Lett.}
  \textbf{\bibinfo{volume}{108}}, \bibinfo{pages}{235302}
  (\bibinfo{year}{2012}).
\end{thebibliography}
\end{document}